\documentclass[aps,prb,twocolumn,amsmath,amssymb,groupedaddress]{revtex4}

\usepackage{dcolumn}
\usepackage{graphicx,subfigure}
\usepackage{bm}
\usepackage{verbatim}
\usepackage{amsmath}
\usepackage{amssymb}
\usepackage[T1]{fontenc}
\usepackage{ae,aecompl}
\usepackage{appendix}
\usepackage{float}
\usepackage{color}

\newcommand{\rv}{{\bf r}}

\newcommand{\av}{{\bf a}}

\newcommand{\xv}{{\bf x}}

\newcommand{\bv}{{\bf b}}
\newcommand{\ev}{{\bf e}}

\newcommand{\jv}{{\bf j}}
\newcommand{\qv}{{\bf q}}

\newcommand{\nv}{{\bf{n}}}

\newcommand{\pv}{{\bf p}}

\newcommand{\oh}{{\frac{1}{2}}}

\newcommand{\grad}{{\bm{\nabla}}}

\newcommand{\be}{\begin{equation}}
\newcommand{\ee}{\end{equation}}
\newcommand{\bea}{\begin{eqnarray}}
\newcommand{\eea}{\end{eqnarray}}
\newcommand{\bse}{\begin{subequations}}
\newcommand{\ese}{\end{subequations}}
\def\rf#1{(\ref{#1})}

\begin{document}
\title{Fractonic gauge theory of smectics}
\author{Zhengzheng Zhai and Leo Radzihovsky}
\affiliation{
Department of Physics and Center for Theory of Quantum Matter\\
University of Colorado, Boulder, CO 80309}
\date{December 8, 2020}
\email{radzihov@colorado.edu, zhzh3530@colorado.edu}

\begin{abstract}
  Motivated by striped correlated quantum matter, and the recently
  developed duality between elasticity of a two-dimensional (2D)
  crystal and a gauge theory, we derive a dual coupled U(1) vector
  gauge theory for a two-dimensional (2D) quantum smectic, where the
  disclination is mapped onto the fractonic charge, that we
  demonstrate can only move transversely to smectic layers. This
  smectic gauge theory dual also emerges from a gauge dual of a
  quantum crystal through a Higgs transition corresponding to a single
  flavor of its dipole condensation, an anisotropic quantum melting
  via dislocation proliferation. A condensation of the second flavor
  of dislocations corresponds to another Higgs transition describing
  the smectic-to-nematic melting. We also utilize the electrostatic
  limit of this duality to formulate a melting of a 2D classical
  smectic in terms of a higher derivative sine-Gordon model,
  demonstrating its instability to a nematic at any nonzero
  temperature. Generalizing this classical duality to a 3D smectic,
  gives a formulation of a 3D nematic-to-smectic transition in terms of
  an anisotropic Abelian-Higgs model.
  
\end{abstract}

\maketitle

\section{Introduction}
\subsection{Motivation and background}
A smectic state of matter, is a liquid crystal phase that partially
breaks rotational and translational symmetries spontaneously,
exhibiting a periodic layered order. Classical smectics form in
systems of rod-like constituents (molecules like 5CB)
\cite{ProstDeGennes,Chaikin2000} and are driven by anisotropic
entropic (exclusion volume) interactions. In striking contrast,
quantum smectic states appear even in systems of isotropic point-like
constituents as a result of frustrated competition between kinetic
energy and interactions. In cold atom systems, quantum smectics may be
realized in a putative Fulde-Ferrell-Larkin-Ovchinnikov paired
superfluids\cite{FF,LO} in imbalanced degenerate atomic gases
\cite{LR_VishwanathPRL,LRpra} and in spin-orbit coupled Bose
condensates \cite{HuiZhai,LR_ChoiPRL}. Quantum smectics are also a
natural explanation for a striking resistive anisotropy observed in
quantum Hall systems at half-filled high Landau levels
\cite{EisensteinQSm,
  CsathyARCMP,Fogler,Moessner,FisherMacdonald,LR_Dorsey}, and for
``striped'' spin and charge states of weakly doped correlated quantum
magnets\cite{TranquadaStripes, KivelsonStripes}.

A two-dimensional (2D) smectic can emerge from partial, anisotropic
melting\cite{HalperinOstlund} of a crystal, with only one species of
dislocations unbinding, such that only one direction of translational
symmetry is restored, in a Kosterlitz-Thouless (KT)-like \cite{KT}
phase transition. However, a 2D smectic is unstable to thermal
fluctuations, and is always driven into a nematic fluid at any nonzero
temperature \cite{HalperinOstlund, Landau37,
  Peierls36,TonerNelson}. In contrast, a (2+1)D quantum smectic at
zero temperature, is a stable state of matter, whose studies have been
limited to a simplest harmonic description, with effects of
topological defects and of elastic nonlinearities neglected beyond
qualitative discussions (for an exception see
Refs.~\onlinecite{LRpra,Grinstein82, LR2011PRE}). This, together with
ubiquitous putative realizations provides a strong motivation for the
present detailed work, a brief preview of which appeared in a recent
publication.\cite{Smecticgauge}

A complementary motivation for our study is its relation to a new
class of topological quantum states of matter -- dubbed ``fractons''
-- discovered in theoretical exactly solvable models.\cite{Chamon05,
  Bravyi11, Haah11, Castelnovo12, Yoshida13, Bravyi13, Vijay15,
  Vijay16} These feature a number of fascinating properties that are
believed to lie beyond a conventional quantum field theoretic
description.\cite{QiAOP2020} The most striking of these are
quasiparticles with robust (not just fine-tuned or symmetry imposed)
restrictions on their mobility, such as an immobile fracton, and its
subdimensional multipoles.  Although experimental realizations have
been sorely lacking, these theoretical models are intensely studied,
motivated by their promise for a robust quantum memory\cite{Haah11}
and fundamental interest in a new class of topological quantum
liquids\cite{HNreview,AbhinavPremReview}.

Following this gapped class of lattice qubit models, fracton-like
phenomena were also uncovered in gapless symmetric tensor gauge
theories, encoded in a generalized Gauss law, that conserves charge
multipoles and thereby constrains mobility of charges\cite{Pretko1703,
  Pretko1707, Slagle}.  Contemporaneously, a similarity of the constrained
dynamics of disclination and dislocation defects in a crystal was
conjectured to be dual to charges and dipoles of a gauge
theory\cite{RadzihovskyConjectureDuality16}. Utilizing a
generalization of the familiar XY-to-gauge theory (boson-vortex)
duality\cite{dasgupta,fisher}, Pretko and
Radzihovsky\cite{PretkoLRdualityPRL2018} formalized this relation
through a duality mapping (explored in other contexts by Zaanen and
company \cite{Zaanen2017}) between a quantum 2D crystal elasticity and
a symmetric tensor gauge theory. Under this mapping the stress tensor
$\sigma_{ij}$ and momentum vector $\pi_i$ fields map onto the electric
tensor $E_{ij}$ and magnetic vector $B_i$ fields, respectively, with
Newton's law (conservation of momentum) corresponding to Faraday's law
of the tensor gauge theory.  Relation of these tensor gauge theories
to chiral topological elasticity was also explored in
Ref.\onlinecite{GromovDualityPRL2019}.

This established fracton-elasticity duality allows for numerous
predictions for phases and phase transitions of the fracton system,
based on the extensive understanding of 2D crystal and their descendent
states. For example, different phases of the scalar fracton model -
fracton insulator, dipole condensate and fracton condensate, can be
regarded as gauge theory counterparts to the ``commensurate'' and
``incommensurate'' (supersolid)
crystals\cite{PretkoLRsymmetryEnrichedPRL2018, PretkoZhaiLRdualityPRB,
  Kumar19}, hexatic, and isotropic fluid phases of the elasticity
theory. The associated finite-temperature dipole-unbinding transition
and fracton charge unbinding transition correspond to the classical
two-stage melting transitions, i.e., crystal-to-hexatic and
hexatic-to-liquid transition respectively.

A complementary and physically more transparent formulation of
elasticity-to-fractonic coupled {\em vector} gauge theory was recently
presented\cite{RadzihovskyHermeleVectorGaugePRL2020}. The resulting
dual coupled vector gauge theory involves three U(1) vector gauge
fields ${\bf A}_k$ (with $k=x,y$ denoting flavors) and $\av$, and
their canonically conjugate electric fields ${\bf E}_k$ and $\ev$,
that encode coupled Goldstone modes, the phonons $u_k$ and the local
bond angle $\theta$. Building on the treatment of the quantum
crystal\cite{RadzihovskyHermeleVectorGaugePRL2020} and a recent
analysis of the quantum smectic\cite{Smecticgauge}, we derive and
explore extensively the coupled vector gauge theory duality to study
the (2+1)D smectic and its quantum phase transitions to a crystal and
a nematic, formulated in terms of an array of Higgs transitions. We
also utilize it to study 2D and 3D classical smectic and the
corresponding classical nematic-to-smectic phase transitions
\cite{deGennes72, HalperinMa, Helfrich78, NelsonToner, Lubensky81,
  Grinstein86, Toner82}.

\subsection{Summary of Results}

In this paper, we develop and explore in detail a dual coupled U(1)
vector gauge theory for a 2D quantum smectic, building on a recent
study of a 2D quantum
crystal\cite{RadzihovskyHermeleVectorGaugePRL2020} and a
smectic\cite{Smecticgauge} by one of the authors.  The dual
description we derive is formulated in terms of two coupled U(1)
vector gauge theories, with electric fields $\hat{{\bf E}}$ and
$\hat{{\bf e}}$, and canonically conjugate vector potentials
$\hat{{\bf A}}$ and $\hat{{\bf a}}$, sourced by dipole and charge
current densities, $n_b, {\bf j}_b$ (dislocations) and $n_s, {\bf j}_s$
(disclinations), respectively. The corresponding dual Hamiltonian
density is given by
\begin{equation}
\begin{split}
    \tilde{\mathcal{H}}_{\text{sm}}=&\frac{1}{2}\chi \hat{\bf E}^2+\frac{1}{2}\left(\grad \times \hat{\bf A}\right)^2+\frac{1}{2}K \hat{\bf e}^2\\&+\frac{1}{2}\left( \grad \times \hat{\av}+\hat{\bf x} \times \hat{\bf A} \right)^2-\hat{\bf A} \cdot {\bf j}_b-\hat{\bf a} \cdot {\bf j}_s,
\end{split}
\label{H_dual}
\end{equation}
supplemented by the generalized Gauss laws,
\begin{eqnarray}
   \grad \cdot \hat{{\bf E}}&=&n_b+\hat{{\bf e}} \cdot \hat{\bf x},\\
   \grad \cdot \hat{{\bf e}}&=&n_s.
\end{eqnarray}

We demonstrate that the $n_s$ charges (disclinations in the smectic)
of this dual gauge theory are subdimensional ``lineon'', mobile only
transverse to smectic layers (that we take to be along $\hat{\bf x}$), enforced by
generalized gauge invariance and associated continuity equation,
\begin{eqnarray}
    \partial_t n_b+\grad \cdot {\bf j}_b&=&-\hat{\bf x}\cdot{\bf j}.
\end{eqnarray}
In contrast the dipoles (dislocations) exhibit a finite but highly
anisotropically mobility.

\begin{figure}[htbp]
\centering
 \hspace{0in}\includegraphics*[width=0.5\textwidth]{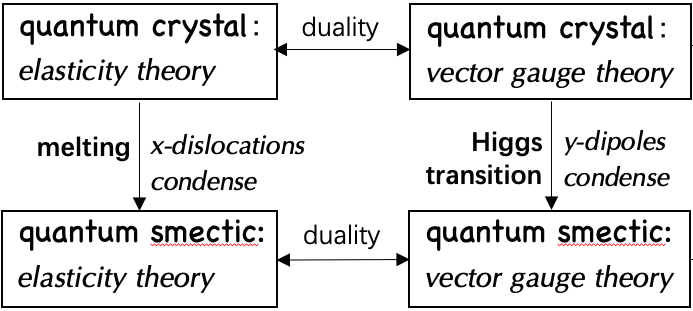}
 \caption{Quantum crystal-smectic duality relations and the associated
   quantum melting transition.}
\label{fig:Crystal-smectic chart}
\end{figure}

Motivated to also understand the quantum crystal-smectic transition,
we derive the smectic gauge dual and transition to it by utilizing
gauge dual of the quantum
crystal\cite{RadzihovskyHermeleVectorGaugePRL2020} and condensing one
flavor of dipoles (dislocations). The associated Higgs transition gaps
out the corresponding flavor of the gauge fields ${\bf A}_k$, and
leads to a dual quantum smectic Lagrangian, that matches exactly the
description obtained through direct duality of smectic elasticity,
\begin{equation}
  \tilde{\mathcal{L}}_{\text{sm}}=\frac{1}{2} |\left(\partial_{\mu}+ipA_{\mu} \right)\psi_x|^2-V\left(|\psi_x|\right)+\mathcal{L}_{\text{M}}^{\text{sm}},
\end{equation}
where, $\mathcal{L}_{\text{M}}^{\text{sm}}$ is the Maxwell sector from Equation (\ref{H_dual}), with $V\left(|\psi_x|\right)$ a U(1)-invariant Landau
potential for $x$-flavor dipoles, $\psi_x$.  The flow chart in
Fig. \ref{fig:Crystal-smectic chart} summarizes these two routes to
the dual gauge theory of a quantum smectic. 

In this formulation the $\psi_x = 0$ Coulomb phase corresponds to
the quantum smectic phase, and the $\psi_x \neq 0$ Higgs phase gives
a condensation of unbound $\hat{\bf x}-$dipoles ($\hat{\bf
  y}-$dislocations) that gaps out the gauge field $A_{\mu}$, which
drives a Higgs transition to a quantum nematic. The quantum melting
transitions and the corresponding phases are illustrated in
Fig. \ref{fig:phase transition}.

 \begin{widetext}
 \begin{center}
 \begin{figure}[htbp]
 \centering
 \hspace{0in}\includegraphics*[width=1.0\textwidth]{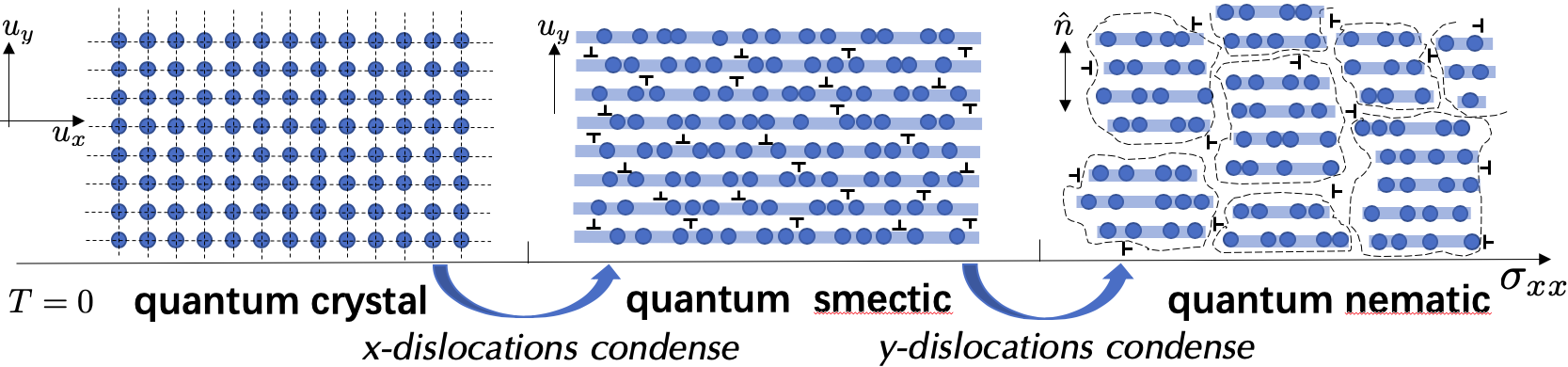}
 \caption{Illustration of quantum melting of a 2D crystal into a
   smectic, followed by smectic-to-nematic melting, respectively
   driven by a condensation of $\hat{\bf x}$-dislocations and of
   $\hat{\bf y}$-dislocations, and tuned by applied shear stress,
   $\sigma_{xx}$.}
\label{fig:phase transition}
\end{figure}
\end{center}
\end{widetext}

\begin{figure}[htbp]
\centering
 \hspace{0in}\includegraphics*[width=0.5\textwidth]{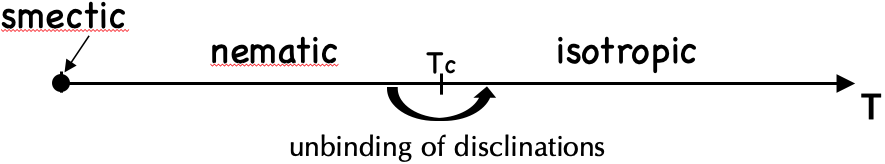}
 \caption{A phase diagram illustrating an instability of a 2D smectic to
   arbitrary weak thermal fluctuations that drive it into a nematic at
   any nonzero temperature. The nematic phase transition into an
   isotropic fluid is through a KT disclination unbinding transition
   $T_c$.}
\label{fig:SmecticMelting}
\end{figure}

We also explore the classical limit of this duality, and formulate the
2D smectic-to-nematic melting and a subsequent nematic-to-isotropic
fluid transition in terms of a higher-derivative sine-Gordon model,
\begin{equation}
\begin{split}
  \tilde{\mathcal{H}}_{\text{sm}}=&\frac{1}{2} \chi^{-1} (\partial_x^2 \alpha)^2+\frac{1}{2} K^{-1} (\partial_y \alpha)^2-g_b \cos (b\partial_x \alpha)\\&-g_s \cos (2\pi \alpha).
\end{split}
\end{equation}
The first two terms capture the elasticity of a 2D smectic, and the
two cosine correspond to dislocations and disclinations, tuned by the
corresponding fugacities $g_{b,s}$. We demonstrate that in 2D the
dislocations are always relevant, corresponding to the instability of
a 2D smectic to a nematic at any nonzero temperature
\cite{HalperinOstlund, Landau37, Peierls36,TonerNelson}.  The
resulting sine-Gordon model in $\alpha$ then captures the the
nematic-to-isotropic fluid transition, illustrated in
Fig. \ref{fig:SmecticMelting}.

We also derive a classical dual gauge theory for a 3D smectic, that
captures its finite-temperature melting into a nematic through a dual
normal-superconductor transition with an higher-derivative Maxwell
sector, equivalent to Toner's original treatment of the
nematic-to-smectic-A transition.\cite{Toner82}

\subsection{Outline}
\label{sec:outline}

The rest of this paper is organized as follows. In Sec. \ref{sec:Duality}, after
briefly introducing the elasticity theory of a smectic phase and its
topological defects, we map a two-dimensional quantum smectic to a
dual coupled U(1) vector gauge theory, and use it to demonstrate that
its charges (disclinations) exhibit subdimensional constrained
mobility. In Sec. \ref{sec:Higgs}, starting with the coupled U(1) vector gauge
theory for a quantum crystal, and ``softening'' it into a generalized
Abelian-Higgs model, we rederive the dual gauge theory of a quantum
smectic through a Higgs transition of one flavor of its
dipoles. Furthermore, we derive an equivalent low-energy tensor
gauge-theory description. In Sec. \ref{sec:Classical limit}, we explore the classical
analogue of these dualities and associated phase transitions, and
formulate a higher derivative sine-Gordon model, capturing classical
thermal smectic melting transitions. We use it to demonstrate that
indeed a 2D smectic is unstable and driven into a nematic at any
nonzero temperature. We also generalize this discussion to a 3D
classical smectic, and reformulate the 3D nematic to smectic-A
transition mediated by unbinding of dislocation loops in terms of a
higher-derivative classical normal-superconductor transition. We
conclude in Sec. \ref{sec:summary} with a summary of our results and
discussion of potential utility of our work.

\section{Smectic and its duality}
\label{sec:Duality}
\subsection{ Classical smectic}
\label{sec:elasticity}
Ideal smectics are equidistantly layered structures, with a
well-defined interlayer spacing $d$, which can be determined through
diffraction experiments. With the layers correlations are liquid-like
and exhibit crystal-like periodic modulation transverse to the layers,
with corresponding density given by,
 \begin{equation}
     \rho({\bf r})=\rho_0+ \left(\psi e^{{\bf q}_0 \cdot {\bf r}}+h.c. \right),
 \end{equation}
 where, ${\bf q}_0=\frac{2\pi}{d} \hat{\bf z}$ is the modulation
 wavevector, and $\psi$ its amplitude, that is the order parameter
 that distinguishes the smectic phase from the nematic phase.
 
\begin{figure}[htbp]
 \centering
  \subfigure[]{
   \hspace{0in} \includegraphics[width=1.3in]{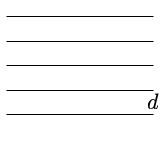}}
    \label{fig:equilibrium}
  \hspace{0.8cm}
  \subfigure[]{
    \hspace{0in}\includegraphics[width=1.4in]{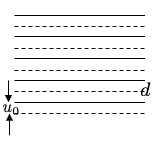}}
    \label{fig:translation}
 \hspace{0.2cm}
  \subfigure[]{
   \hspace{0in} \includegraphics[width=1.3in]{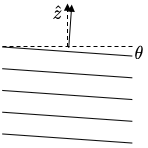}}
    \label{fig:rotation}
  \hspace{0.7cm}
  \subfigure[]{
   \hspace{0in} \includegraphics[width=1.4in]{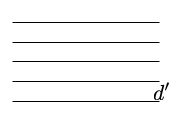}}
    \label{fig:compression}
    \caption{Elastic deformations of a smectic. (a) Smectic with
      equilibrium layer spacing: $q_0=\frac{2\pi}{d}, u=0$. (b) A
      uniformly translated smectic with the same energy as that in
      (a): $u=u_0$. (c) A rotated smectic with the same energy as that
      in (a): with $u=\theta x$ to harmonic order. (d) A smectic with
      compressed layers with energy increased relative that in (a):
      $q_0'=\frac{2\pi}{d'}, u=\left(1-\frac{d}{d'}\right)z$. [Figures
      adapted from Reference \onlinecite{Chaikin2000}.]}
\label{fig:elastic energy}
\end{figure}

 The deformation of a smectic can be described by its layer
 displacement field $u({\bf r})$. As the system is invariant under
 uniform translations, the elastic energy should be expressed purely
 in terms of derivatives of $u$. Furthermore, the first-order
 derivatives along the layer, $\grad_{\perp} u$, corresponding to
 merely a uniform rotation of the layers, must cost no energy. Thus,
 to harmonic order, only the curvature of the layers,
 $\grad_{\perp}^2 u$, can enter the quadratic part of the elastic
 energy functional. This point can be seen more explicitly by the
 following argument. The order parameter $\psi({\bf r})$, describing
 phonon fluctuations, can be represented as
 \begin{equation}
     \psi({\bf r}) = |\psi| e^{-i q_0 u({\bf r})},
 \end{equation}
 and the locations of the layer planes can be determined as the constant phase of the molecular density wave,
 \begin{equation}
 \label{layer locations}
     \phi({\bf r}) \equiv {\bf q}_0 \cdot {\bf r}-q_0 u({\bf r})=2\pi n, n=0, \pm 1, \pm 2, ...
 \end{equation}
The layers local unit-normal is given by,
 \begin{equation}
 \begin{split}
 \label{Normal}
     {\bf N}=& \frac{\grad \phi}{|\grad
       \phi|}=\frac{\left(-\grad_{\perp}u, -\nabla_{\parallel}u, 1\right)}{\sqrt{1+\left(\grad u\right)^2}}\\=& \left(-\partial_x u, -\partial_y u, 1 \right)+O[\left(\grad u\right)^2].
\end{split}
 \end{equation}
 The first-order derivative, $\grad_{\perp}u \approx \left(\partial_x
   u, \partial_y u, 0 \right) \approx \hat{\bf z}-{\bf N}$, therefore,
 corresponds to a rigid rotation of the layers around an axis along
 the layer plane, and does not contribute to the elastic energy, and,
 $\nabla_{\parallel}u=\partial_z u$, in 3D. (See. Fig.\ref{fig:elastic energy}.)
 
Consistent with this, the continuum elastic Hamiltonian density for a
D-dimensional smectic is given by a well-known expression,
 \begin{equation}
 \label{H_sm, d}
      \mathcal{H}_{\text{sm}}=\frac{1}{2}\chi \left(\nabla_{\parallel} u \right)^2+\frac{1}{2}K \left(\grad_{\perp}^2 u \right)^2,
 \end{equation}
 a Landau-Peierls elastic energy \cite{Landau37, Peierls36, LR2011PRE}
 for a one-dimensional solid, where $\chi$ is inverse of the
 compressional modulus, and $K$ is the bend modulus.

Note that \rf{Normal} is only correct up to $O[(\grad u)^2]$ for small rotations. For any finite rotations $\theta$,
 \begin{equation}
     \nabla_{\parallel} u =1-\cos \theta,\;\; \nabla_{\perp} u= -\sin \theta,
 \end{equation}
 and the nonlinear strain, $\nabla_{\parallel} u-\frac{1}{2} (\grad
 u)^2$, can be straightforwardly seen to be independent of the rotation angle
 $\theta$. Thus, the rotationally invariant energy density is given by,
 \begin{equation}
      \mathcal{H}_{\text{sm}}=\frac{1}{2}\chi \left[\nabla_{\parallel} u-\frac{1}{2} (\grad u)^2 \right]^2+\frac{1}{2}K \left(\grad_{\perp}^2 u \right)^2,
 \end{equation}
 which introduces non-linear elasticity into
 $\mathcal{H}_{\text{sm}}$, that for $d\leq 3$ leads to a nontrivial
 anomalous smectic elasticity \cite{Grinstein82, LR2011PRE}. However,
 because the focus of our work is on a quantum smectic, these elastic
 nonlinearities remain irrelevant in (2+1)D and will thus be neglected
 in the rest of the manuscript.
 
 For the smectic-A phase, the local normal field (layer orientation)
 ${\bf N}$ and the director field ${\bf n}=\hat{\bf z}+\delta {\bf n}$
 are aligned in equilibrium. Thus,
 \begin{equation}
     \grad_{\perp} u=- \delta {\bf n},\;\; \grad_{\perp}^2 u=-\grad_{\perp} \delta {\bf n}=-\grad {\bf n},
 \end{equation}
 and the elastic energy, ignoring nonlinearities, can be represented as
 \begin{equation}
     \mathcal{H}_{\text{sm}}=\frac{1}{2}\chi \left(\grad u+ \delta {\bf n} \right)^2+\frac{1}{2}K \left(\grad {\bf n} \right)^2.
 \end{equation}
 For a 2D smectic, with the layers along $\hat{\bf x}$ (with layer normal
 along $\hat{\bf y}$), the elastic Hamiltonian density in Eq.\rf{H_sm,
   d} reduces to
\begin{equation}
    \mathcal{H}_{\text{sm}}=\frac{1}{2}\chi \left(\partial_y u \right)^2+\frac{1}{2}K \left(\partial_x^2 u \right)^2,
 \label{H_el}
\end{equation}
where the layer displacement $u$ is along the $\hat{\bf y}$ axis, and the layer orientation (director field) is, ${\bf n}=-\hat{\bf x} \sin \theta+\hat{\bf y} \cos \theta=\hat{\bf y}+\delta {\bf n}$.

Another way to obtain smectic elasticity is to start out with a
elasticity of a 2D crystal and allow nonsingle-valued displacement
field $u_x$, with $\grad u_x={\bf v}_x$, accounting for a plasma of
unbound dislocations with Burgers vector along the $\hat{\bf
  x}-$directed smectic layers, where ${\bf v}_x$ is an arbitrary
vector with $\epsilon_{ij}\partial_i v_{jx} = b_x$. Integrating over
strain tensor field ${\bf v}_x$, leads to the smectic harmonic
elasticity,
\begin{equation}
\begin{split}
    \mathcal{H}_{\text{sm}}=&\mathcal{H}_{\text{cr+disl.}}=\mu u_{ij}^2+\frac{1}{2}\lambda u_{ii}^2\\
    =&\mu \left(u_{xx}^2+u_{yy}^2+2 u_{xy}^2 \right)+\frac{1}{2} \lambda \left(u_{xx}+u_{yy} \right)^2+\frac{1}{2}E_c b_x^2\\
    =&\mu \left[v_{xx}^2+\left(\partial_y u_y\right)^2+\frac{1}{2} \left(\partial_x u_y+v_{yx} \right)^2 \right]\\&+\frac{1}{2} \lambda \left(v_{xx}+u_{yy} \right)^2+\frac{1}{2}E_c \left(\partial_x v_{yx}-\partial_y v_{xx}\right)^2\\=& \frac{1}{2} \chi \left(\partial_y u_y \right)^2+\frac{1}{2}K \left(\partial_x^2 u_y \right)^2,
\end{split}
\label{H_cr+disl}
\end{equation}
where compressional modulus is $\chi = 4\mu (\mu + \lambda)/(2\mu +
\lambda)$, bend modulus $K = E_c$, and higher derivative terms are
neglected after integrating out the ${\bf v}_x$ field in the last
step.

Equivalently, the smectic elasticity can be formulated in terms of the
orientational (nematic) angle degree of freedom $\theta$, which
corresponds to the orientation of the layers, with the elastic
Hamiltonian density given by,
\begin{equation}
    \mathcal{H}_{\text{sm}}=\frac{1}{2} \chi \left(\grad u -\theta \hat{\bf x}\right)^2+\frac{1}{2} K \left(\grad \theta \right)^2.
\label{H_el_u_theta}
\end{equation}
At low energies set by $\chi$, this Hamiltonian reduces to the
conventional form (\ref{H_el}) after Higgs'ing out the bond angle
$\theta$, locking $\partial_xu=\theta$.

\subsection{Two-dimensional quantum smectic}
\label{sec: 2D quantum smectic}
Classical elastic Hamiltonian in (\ref{H_el_u_theta}) is easily
generalized to a quantum smectic by elevating $u$ and $\theta$ to
operators, and adding canonically conjugate linear and angular momenta
operators, $\hat{\pi}$ and $\hat{L}$, respectively. This gives,
\begin{equation}
    \hat{\mathcal{H}}_{\text{sm}}=\frac{1}{2} \hat{\bf \pi}^2 +\frac{1}{2}\hat{L}^2+\frac{1}{2} \chi \left(\grad \hat{u} -\hat{\theta} \hat{\bf x}\right)^2+\frac{1}{2} K \left(\grad \hat{\theta} \right)^2,
\end{equation}
for bosonic smectic supplemented with canonical commutation relations
($\hbar=1$),
\begin{subequations}
\begin{eqnarray}
    \left[\hat{u}({\bf r}), \hat{\pi} ({\bf r}') \right]&=&i \delta^2 \left({\bf r}-{\bf r}' \right),\\
    \left[\hat{\theta} ({\bf r}), \hat{L} ({\bf r}') \right]&=&i \delta^2 \left({\bf r}-{\bf r}' \right).
\end{eqnarray}
\end{subequations}

It is convenient to work with a path-integral formulation where
quantum nature of these fields is accounted for by functional
integration in phase-space of these fields. We consider the evolution
operator for the quantum smectic,
\begin{equation}
U(u', \theta', u, \theta;t)=\langle u', \theta'| e^{-i\int_{\rv} \hat{\mathcal{H}}_{\text{sm}}t}|u, \theta\rangle,
\end{equation}
and rewrite it in phase-space functional integral formulation as,
\begin{equation}
  \begin{split}
   U&=\int \left[du\right] \left[d\pi\right] \left[d\theta\right] \left[dL\right] e^{iS_{\text{sm}}}\\
   &=\int \left[du\right] \left[d\pi\right] \left[d\theta\right] \left[dL\right] e^{i\int dt \int_{\rv} \mathcal{L}_{\text{sm}}},
  \end{split}
\end{equation}
with the corresponding Lagrangian density given by,
\begin{equation}
\begin{split}
    \mathcal{L}_{\text{sm}}=&\pi \partial_t u+L\partial_t \theta-\frac{1}{2}\pi^2-\frac{1}{2}L^2+\frac{1}{2}\chi^{-1} {\bf \sigma}^2+\frac{1}{2} K^{-1} {\jv}^2 \\&-{\bf \sigma} \cdot \left(\grad u -\theta \hat{\bf x}\right)-{\jv} \cdot \grad\theta.
\end{split}
\end{equation}

\begin{figure}[htbp]
  \centering
  \subfigure[]{
    \hspace{0in}\includegraphics[width=0.22\textwidth]{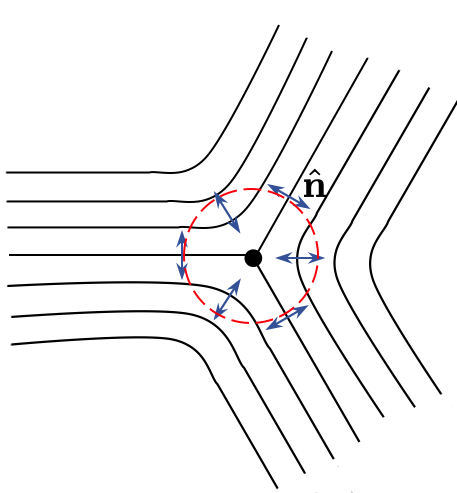}}
    \label{fig:-disclination}
  \hspace{0.3cm}
  \subfigure[]{
    \hspace{0in}\includegraphics[width=0.22\textwidth]{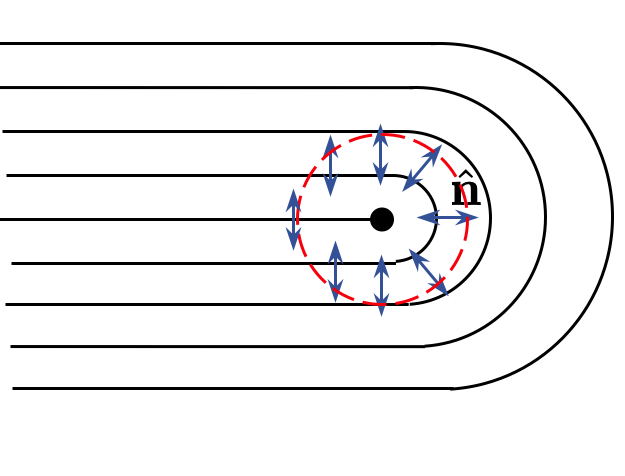}}
    \label{fig:+disclination}
  \hspace{0.4cm}
  \subfigure[]{
    \hspace{0in}\includegraphics[width=0.33\textwidth]{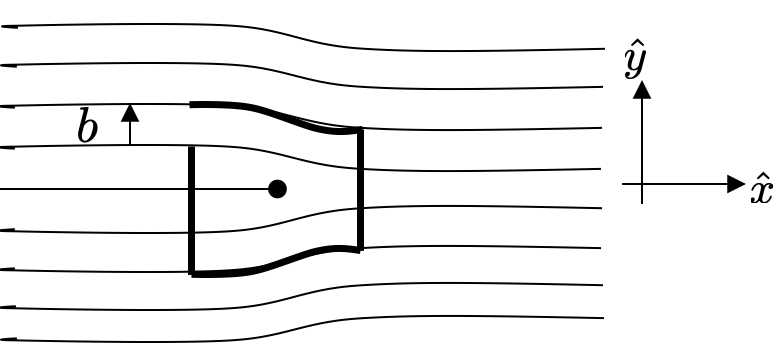}}
    \label{fig:dislocation}
    \caption{Topological defects in a 2D smectic. $\hat{\nv}$ is the Frank director, and $\bv$ is the Burger's vector. (a) A $-\frac{1}{2}$
      disclination,  i.e., the director $\hat{\nv}$ changes $-\pi$ around a closed loop counterclockwise. (b) A $+\frac{1}{2}$
      disclination, i.e., the director $\hat{\nv}$ changes $+\pi$ around a closed loop counterclockwise. (c) A single positive dislocation, i.e., the layer displacement $u$ changes $+d$ around a closed circuit counterclockwise, which can be regarded as  a tightly bound dipole of two, opposite disclinations with charges $\pm \frac{1}{2}$.}
\label{fig:defects}
\end{figure}

In addition to the single-valued (smooth) Goldstone mode degrees of
freedom, $\tilde{\theta}$ and $\tilde{u}$, we must also include
topological defects -- disclinations and dislocations, captured by
including a nonsingle-valued component of the bond angle field $\theta^s$
and of the phonon distortion field $u^s$, respectively. In the
smectic, a disclination at a point ${\bf r}_0$, is defined by a
nonzero closed line-integral of the gradient of the bond angle around
${\bf r}_0$, $\oint_{\rv_0} d\theta=2\pi n_s$, or equivalently in a
differential form,
\begin{equation}
    \hat{\bf
      z}\cdot\grad\times\grad\theta^s=2\pi n_s\delta^2({\bf
      r}-\rv_0)\equiv n_s({\bf r}),
\label{defn_s}
\end{equation}
measuring the deficit/surplus bond angle. $n_s({\bf r})$ is the
disclination charge density. A disclination with charge
$n_s=-\frac{1}{2}$, and a disclination with charge $n_s=+\frac{1}{2}$
are illustrated in Fig. \ref{fig:defects}(a, b).

A dislocation at ${\bf r}_0$ with a Burgers charge $n_b$ (that is an
integer multiples of the elementary layer spacing), is defined by a
closed line-integral, $\oint_{\rv_0} d u = n_b$, or equivalently in
the differential form,
\begin{equation}
  \hat{\bf z}\cdot\grad\times\grad u^s=n_b\delta^2({\bf
    r}-\rv_0)\equiv n_b({\bf r}),
\label{defn_b}
\end{equation}
where $n_b({\bf r})$ is the Burgers charge density. A dislocation in
the smectic is shown in Fig. \ref{fig:defects}(c), which can be regarded as a tightly
bound pair of $+\frac{1}{2}$ and $-\frac{1}{2}$ disclinations.

In anticipation of our more rigorous duality derivation, already here
we can argue for the subdimensional nature of the disclination
dynamics in a smectic. Consider a pair of oppositely charged $\pm
\frac{1}{2}$ disclinations separated along $\hat{x}$ axis, as shown in
Fig. \ref{fig:disclination mobility}. The separation between the pair
is large such that we can regard them as isolated
disclinations. Moving the `$+$' disclination along the layers (i.e.,
$\hat{\bf x}$ axis) by two layer spacings, requires the introduction
of four extra half-layers of molecules, which is a highly non-local
process in terms of atoms quantum dynamics, and is therefore not
allowed.  While moving the `$+$' disclination transversely (i.e., in
$\hat{y}$ direction) preserves the strength of the dislocation, and
thus, is allowed dynamically, even though there is an energy cost for
separating the `$+$' and `$-$' disclination pair that make up the
dislocation. Similar analysis applies to the `$-$' disclination. Thus,
we conclude that disclinations can only move transversely to the
layers, i.e., they manifests the subdimensional lineon dynamics.

To include the topological defects in the complete description of the
smectic, we decompose the distortion field $u$ and the bond angle
$\theta$ into the smooth elastic and nonsingle-valued components,
\begin{equation}
    u=\tilde{u}+u^s,\ \ \theta= \tilde{\theta}+\theta^s.
\end{equation}
Integrating out the single-valued parts $\tilde{u}$ and
$\tilde{\theta}$ out of the total generating function,
\begin{equation}
   U=\int \left[d\tilde{u}\right] [d\tilde{\theta}] \left[d u^s\right] \left[d\pi\right]  \left[d\theta^s\right] \left[dL\right] e^{i\int dt \int_{\rv} \mathcal{L}_{\text{sm}}},
\end{equation}
leads to,
\begin{equation}
    U=\int \left[du^s\right] \left[d\pi\right] \left[d\theta^s\right] \left[dL\right] e^{i\int dt \int_{\rv} \mathcal{L}_{\text{sm}}},
\end{equation}
where the new Lagrangian density is given in terms of only nonsingular components,
\begin{equation}
  \begin{split}
    \mathcal{L}_{\text{sm}}=&\pi \partial_t u^s+L\partial_t \theta^s-\frac{1}{2}\pi^2-\frac{1}{2}L^2+\frac{1}{2}\chi^{-1} {\bf \sigma}^2 \\&-{\bf \sigma} \cdot \left(\grad u^s-\theta^s \hat{\bf x}\right)+\frac{1}{2} K^{-1} {\jv}^2 -{\jv} \cdot \grad\theta^s,
\end{split}  
\end{equation}
with two enforced constraints,
\begin{eqnarray}
    &&\partial_t \pi-\grad \cdot {\bf \sigma}\equiv \partial_{\mu} J_{\mu}=0,\label{momentum continuity}\\
    &&\partial_t L-\grad \cdot \jv-\sigma_x=0,
    \label{dipolecontinuity}
\end{eqnarray}
where we implicitly introduced currents $J_{\mu}=\left(\pi, -\sigma_i \right)$.

\begin{widetext}
\begin{center}
\begin{figure}[htbp]
\centering
  \subfigure[]{
   \hspace{0in} \includegraphics[width=0.9\textwidth]{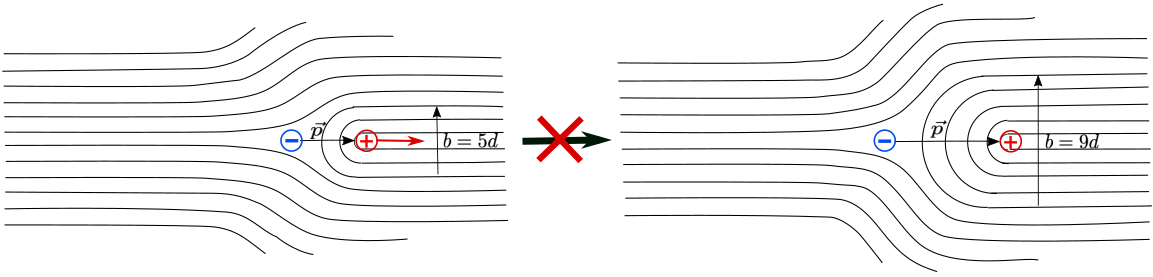}}
    \label{fig:disclinationmobility-alonglayers}
  \hspace{0.15cm}
  \subfigure[]{
   \hspace{0in} \includegraphics[width=0.9\textwidth]{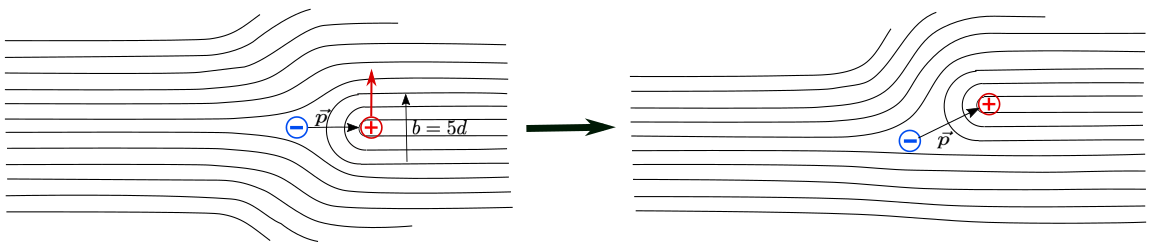}}
    \label{fig:disclinationmobility-transverse}
    \caption{Restricted along-layers mobility of disclinations in a 2D
      quantum smectic. Consider a pair of disclinations separated by
      five layer spacings. (a) Moving the `$+$' disclination along the
      layers by two layer spacings requires an introduction of four
      half-layers of underlying bosons, which is a non-local process,
      and therefore is not allowed. (b) Moving the `$+$' disclination
      transverse to the layers by two layer spacings preserves the
      strength of the dislocation, and thus, is dynamically allowed.}
\label{fig:disclination mobility}
\end{figure}
\end{center}
\end{widetext}

\subsection{Quantum smectic-gauge theory duality}
The momentum continuity equation (\ref{momentum continuity}) can be
solved in terms of gauge potential fields, $A_{\mu}$, with
\begin{equation}
    J_{\mu}=\epsilon_{\mu\nu\gamma}\partial_{\nu} A_{\gamma}=\left(\epsilon_{ij}\partial_i A_j, \epsilon_{i\nu\gamma}\partial_{\nu}A_{\gamma}\right),
\end{equation}
such that
\begin{eqnarray}
    &&\pi=\epsilon_{ij}\partial_i A_j=\hat{\bf z} \cdot \left( \grad \times {\bf A}\right),\\
    &&\sigma_i=-\epsilon_{i\nu\gamma}\partial_{\nu}A_{\gamma}=\epsilon_{ij}(\partial_t A_j-\partial_j A_0),
\end{eqnarray}
and Eq.(\ref{dipolecontinuity}) transforms into 
\begin{equation}
    \partial_t \left(L-\epsilon_{xj}A_j\right)-\partial_i \left(j_i-\epsilon_{xi} A_0 \right) \equiv \partial_{\mu}\tilde{j}_{\mu} =0,
\end{equation}
with $\tilde{j}_{\mu}=\left(L-\epsilon_{xj}A_j,-j_i+\epsilon_{xi}A_0 \right)$, which is then solved by introducing another vector gauge potential $a_{\mu}$,
\begin{equation}
    \tilde{j}_{\mu}=\epsilon_{\mu\nu\gamma}\partial_{\nu}a_{\gamma},
\end{equation}
such that
\begin{eqnarray}
   L&=&\epsilon_{ij}\partial_i a_j+\epsilon_{xj}A_j=\hat{\bf z} \cdot \left(\grad \times {\bf a}+\hat{\bf x} \times {\bf A} \right),\\
   j_i&=&\epsilon_{ij}(\partial_t a_j-\partial_j a_0)+\epsilon_{xi}A_0.
\end{eqnarray}
Substituting the solutions of $\pi, \sigma_i, L$,and $j_i$ in terms of the gauge fields into the Lagrangian density, leads to the dual Lagrangian density
\begin{equation}
     \tilde{\mathcal{L}}_{\text{sm}}=\mathcal{L}_{\text{M}}^{\text{sm}} +{\bf A} \cdot {\bf j}^b+A_0 n_b+{\bf a} \cdot {\bf j}^s+a_0 n_s,
\label{Dual smectic}
\end{equation}
where $\mathcal{L}_{\text{M}}$ is the Maxwell part, given by
\begin{equation}
\begin{split}
    \mathcal{L}_{\text{M}}^{\text{sm}}=&\frac{1}{2} \chi^{-1} \left( \partial_t {\bf A}-\grad A_0\right)^2-\frac{1}{2}\left(\grad \times {\bf A} \right)^2\\&+\frac{1}{2} K^{-1} \left(\partial_t {\bf a}-\grad a_0-A_0 \hat{\bf x}\right)^2-\frac{1}{2}\left(\grad \times {\bf a}+\hat{\bf x}\times {\bf A}\right)^2,
\label{Maxwellsmectic}
\end{split}
\end{equation}
and the charge contributions are obtained by integrating by parts and
defining the dislocation and disclination charge and current densities
as,
\begin{subequations}
\begin{eqnarray}
   n_b&=&\epsilon_{ij}\partial_i\partial_j u,\\
   n_s&=&\epsilon_{ij}\partial_i\partial_j \theta,\\
   j^b_i&=&\epsilon_{ij} \left(\partial_j\partial_t u-\partial_t\partial_j u \right),\\
   j^s_i&=&\epsilon_{ij} \left(\partial_j\partial_t \theta-\partial_t\partial_j \theta \right).
\end{eqnarray}
\end{subequations}
Introducing Hubbard-Stratonovich fields ${\bf E}$ and ${\bf e}$, the
Lagrangian density transforms into
\begin{equation}
\begin{split}
    \tilde{\mathcal{L}}_{\text{sm}}=&-{\bf E} \cdot \left( \partial_t {\bf A}-\grad A_0\right)-\frac{1}{2} \chi {\bf E}^2-\frac{1}{2}\left(\grad \times {\bf A} \right)^2-\frac{1}{2} K {\bf e}^2\\&- {\bf e} \cdot \left(\partial_t {\bf a}-\grad a_0-A_0 \hat{\bf x}\right)-\frac{1}{2}\left(\grad \times {\bf a}+\hat{\bf x}\times {\bf A}\right)^2\\&+{\bf A} \cdot {\bf j}^b+A_0 n_b+{\bf a} \cdot {\bf j}^s+a_0 n_s.
\end{split}
\end{equation}
Integrating over $A_0$ and $a_0$ gives the Gauss law, leaving the
standard Lagrangian form, $\tilde{\mathcal{L}}_{\text{sm}}=-{\bf E}
\cdot\partial_t {\bf A}-{\bf e} \cdot\partial_t {\bf
  a}-\tilde{\mathcal{H}}_{\text{sm}}$, from which we can read off the
dual Hamiltonian density, that is given by,
\begin{equation}
\begin{split}
    \tilde{\mathcal{H}}_{\text{sm}}=&\frac{1}{2}\chi {\hat{\bf E}}^2+\frac{1}{2}\left(\grad \times {\hat{\bf A}} \right)^2+\frac{1}{2}K{\hat{\bf e}}^2\\&+\frac{1}{2}\left(\grad \times {\hat{\av}}+\hat{\bf x} \times {\hat{\bf A}} \right)^2-{\hat{\bf A}} \cdot {\bf j}^b-\hat{\bf a}\cdot {\bf j}^s,
\end{split}
\end{equation}
supplemented by the generalized Gauss laws
\begin{eqnarray}
   \grad \cdot \hat{\bf E}&=&n_b+\hat{\bf e} \cdot \hat{\bf x},\\
   \grad \cdot \hat{\bf e}&=&n_s,
\end{eqnarray}
where, ${\bf E}$ and ${\bf e}$, are independent electric fields, canonically conjugate to the corresponding vector potentials, ${\bf A}$ and ${\bf a}$, respectively.

The above Hamiltonian must be invariant under the gauge transformations:
\begin{subequations}
\begin{eqnarray}
   &&(A_0, A_i) \to A_{\mu}'=\left(A_0+\partial_t \chi, (A_i+\partial_i \chi)\right),\\
   &&(a_0,a_i) \to a_{\mu}'=\left(a_0+\partial_t \phi,(a_i+\partial_i\phi+\hat{\bf x}_i \chi) \right).\;\;\;\;\;\;\;
\end{eqnarray}
\label{gauge transform}
\end{subequations}
 Requiring the source term to preserve this gauge invariance, we obtain coupled continuity equations for charges (disclinations) and dipoles (dislocations), satisfying,
\begin{eqnarray}
    \partial_t n_s+\grad \cdot {\bf j}_s &=& 0,\\
    \partial_t n_b+\grad \cdot {\bf j}_b&=&-j_s^x.
\end{eqnarray}
We observe that the dipole (dislocation) continuity equation is
violated by a nonzero charge (disclination) current $j_s^x$ in the
$\hat{\bf x}$ (along the layers) direction. Thus, in the absence of
gapped $\hat{\bf x}-$dipoles ($\hat{\bf y}$-dislocations), we find
that $j_s^x=0$, i.e., motion of isolated fracton charges
(disclinations) is restricted to be transverse to the smectic layers,
as moving along the layers requires $\hat{\bf x}$-dipoles ($\hat{\bf
  y}$-dislocations) that are gapped in the smectic. Therefore, the
fractons (disclinations) exhibit subdimensional lineon dynamics, as
argued in Section \ref{sec: 2D quantum smectic}.

\section{Higgs transition of quantum crystal-to-smectic melting}
\label{sec:Higgs}
As discussed in the Introduction, a smectic can emerge from
anisotropic melting \cite{HalperinOstlund} of a crystal, understood in
terms of a Kosterlitz-Thouless (KT)-like, \cite{KT} single-species
dislocation unbinding transition. This classical partial melting
transition of an anisotropic solid was studied by Halperin and Ostlund
\cite{HalperinOstlund}, and we have used this fact in
Eq.(\ref{H_cr+disl}) to derive the smectic harmonic elasticity by
including $\hat{\xv}-$orientated dislocations in a classical
crystal. Although such a 2D classical smectic is unstable to thermal
fluctuations, driven into a nematic fluid at any nonzero temperature
\cite{Landau37, Peierls36,HalperinOstlund,TonerNelson} (see Section
\ref{sec:Classical limit}), a (2+1)D quantum smectic at zero
temperature is a stable state of matter. In this section, we
demonstrate that similarly, a quantum smectic can also emerge from
partial, anisotropic quantum melting of a crystal.

The Hamiltonian density of a 2D quantum crystal is,
\begin{equation}
\begin{split}
\mathcal{H}_{\text{cr}}=&\frac{1}{2} C_{ij,kl} (\partial_i \hat{u}_j-\hat{\theta} \epsilon_{ij}) (\partial_k \hat{u}_l-\hat{\theta} \epsilon_{kl})+\frac{1}{2} K (\grad \hat{\theta})^2\\&+\frac{1}{2} \hat{\bf \pi}^2 +\frac{1}{2}\hat{L}^2,
\end{split}
\end{equation}
where, $\hat{\bf u}$ is the phonon field operator, $\hat{\theta}$ is the orientational bond-angle field operator, and, $\hat{\pi}$ and $\hat{L}$ are their corresponding canonically conjugate momentum respectively. $C_{ij,kl}$ is the elastic constant tensor, which takes the form, $C_{ij,kl}=\lambda \delta_{ij}\delta_{kl}+ 2\mu\delta_{ik}\delta_{jl}$, for an isotropic hexagonal lattice, characterized by two independent Lam\'e coefficients, $\lambda$ and $\mu$. Working in the path-integral formulation with the field operators replaced by corresponding classical fields, the action is
\begin{align}
S_{\text{cr}} =& \int dt \int_{\bf r} \frac{1}{2}\left[(\partial_t {\bf u})^2 + (\partial_t\theta)^2 - K (\grad\theta)^2 \nonumber\right.\\& \left.
- C_{ij,kl}(\partial_iu_j - \theta\epsilon_{ij})(\partial_ku_\ell -\theta \epsilon_{kl})  \right].
\label{altelast}
\end{align}

Now, as the $\hat{\bf x}-$dislocations condensed, we include dislocations with Burgers vector along the layers ($\hat{\bf x}$) by replacing $\partial_i u_x = v_{ix}$, where $v_i$ is an arbitrary vector field with $\epsilon_{ij}\partial_i v_{jx} = b_x$. Then, the Lagrangian density
\begin{equation}
\begin{aligned}
\mathcal{L}_{\text{cr+disl.}} =& \frac{1}{2}\left[(\partial_t {\bf u})^2 + (\partial_t\theta)^2 - K(\grad\theta)^2 -\lambda \left(\grad \cdot {\bf u}\right)^2\right.\\&\left.-2\mu (\partial_iu_j - \theta\epsilon_{ij})^2 \right]\\=&\frac{1}{2}\left[(\partial_t {\bf u})^2 + (\partial_t\theta)^2 - K(\grad\theta)^2 -\lambda (v_{xx}+\partial_yu_y)^2\right.\\&\left.-2\mu \left(v_{xx}^2+(\partial_y u_y)^2+(v_{yx}-\theta)^2+(\partial_x u_y-\theta)^2 \right) \right.\\&\left.-\frac{1}{2} E_c (\partial_x v_{yx}-\partial_y v_{xx})^2\right]\\=& \frac{1}{2}\left[(\partial_t {\bf u})^2 + (\partial_t\theta)^2 -K (\grad\theta)^2 - \chi_1 (\partial_y u_y)^2 \right.\\&\left.-\chi_2 (\partial_x u_y-\theta)^2\right]\\=& \frac{1}{2}\left[(\partial_t u_y)^2 + (\partial_t\theta)^2 -\chi \left(\grad u_y-\theta \hat{\bf x}\right)^2-K (\grad\theta)^2\right],
\label{altelast2}
\end{aligned}
\end{equation}
where, in the third line, we have integrated out $v_{ix}$ and $u_x$,
and have defined $\chi_1=4\mu (\mu + \lambda)/(2\mu + \lambda)$ and
$\chi_2=2\mu$, which are taken to be equal in the last line,
$\chi_1=\chi_2=\chi$, for simplicity, corresponding to the case when
$\lambda \to 0$. Then, we arrive at the Lagrangian density of a
smectic, starting with that of a quantum crystal,
\begin{equation}
    \mathcal{L}_{\text{sm}}=\mathcal{L}_{\text{cr+disl.}},
\end{equation}
as summarized by the flow chart in Fig. \ref{fig:Crystal-smectic chart}.

Motivated by this possibility of partial quantum melting of a crystal
into a smectic and the subsequent melting into a quantum nematic, we
will explore its dual in this section. Below, we will also derive
a dual gauge theory of a quantum smectic, through a Higgs transition
from a dual gauge theory of an incommensurate quantum crystal
(supersolid) by condensing one flavor of dipoles, and thereby
Higgs'ing out a flavor component of the gauge fields. As required by
consistency, we indeed find that the resulting quantum smectic dual is
in full agreement with a direct duality derived in Section \ref{sec:Duality}.

\subsection{Soft-spin descriptions of quantum crystal and quantum smectic}
\subsubsection{Quantum crystal}
The dual coupled U(1) vector gauge theory for a quantum crystal was first derived in Ref. [\onlinecite{RadzihovskyHermeleVectorGaugePRL2020}], characterized by the Lagrangian density
\begin{equation}
    \begin{split}
        \tilde{\mathcal{L}}_{\text{cr}}=&\mathcal{L}_{\text{M}}^{\text{cr}}+{\bf A}_k \cdot {\bf J}_k+A_{0k} n_k^b+{\bf a} \cdot {\bf j}^s+a_0 n_s,
    \end{split}
\end{equation}
where $k=(x,y)$ indexes different flavors, ${\bf A}_k$, ${\bf a}$ gauge fields capture the $k=x, y$ phonons and bond orientational order respectively. The Maxwell part, $\mathcal{L}_{\text{M}}^{\text{cr}}$, is given by 
\begin{equation}
   \begin{split}
    \mathcal{L}_{\text{M}}^{\text{cr}}=&\frac{1}{2} \chi^{-1} \left( \partial_t {\bf A}_k-\grad A_{0k}\right)^2-\frac{1}{2}\left(\grad \times {\bf A}_k \right)^2\\+&\frac{1}{2} K^{-1} (\partial_t {\bf a}-\grad a_0- A_{0k} \hat{\bf e}_k)^2-\frac{1}{2}\left(\grad \times {\bf a}-\hat{\bf z}\times {\bf A}_k\right)^2.
\label{Maxwellcrystal}
\end{split} 
\end{equation}

To access descendant phases and corresponding quantum phase transitions, we need to treat dislocation and disclination defects as dynamical charges. Following a standard analysis and focusing on dislocations (dipoles) for the moment, we introduce the dynamical field $\psi_k({\bf r}, t)=\sqrt{\rho_k} e^{i\varphi_k}$ for each gauge-charged dipole species ${\bf p}_k$, and add corresponding kinetic energies, $\frac{\rho_k}{2} \left(\partial_t \varphi_k+pA_{0k}\right)^2$, where we have integrated out the massive magnitude fluctuations to focus on the low-energy phase fluctuations only.
\begin{figure}[htbp]
\label{fig:Dislocationclimb}
\centering
  \subfigure[]{
    \includegraphics[width=0.19\textwidth]{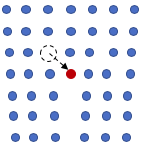}}
    \label{fig:positive-dislocation-climb}
  \hspace{0.8cm}  
  \subfigure[]{
    \includegraphics[width=0.19\textwidth]{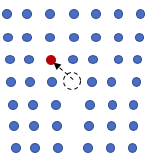}}
    \label{fig:negative-dislocation-climb}
  \caption{Dislocation climb via vacancies diffusion. An edge
    dislocation moves out of the slip plane onto a parallel plane
    directly above or below the slip plane. This movement (climb) is termed nonconservative, as compared with conservative movement (glide). (a) Diffusion of vacancy to edge dislocation. (b) Dislocation climbs up one lattice spacing.}
\label{fig: Dislocation climb}
\end{figure}

\begin{figure}[htbp]
\centering \hspace{0in}\includegraphics*[width=0.28\textwidth]{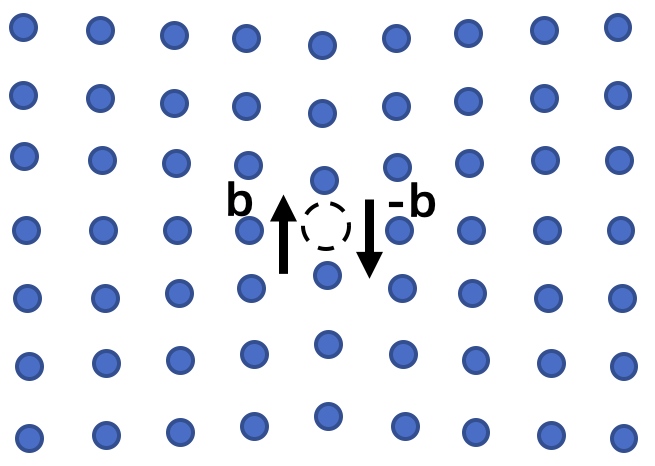}
\caption{A disclination quadrupole, constructed as a bound pair of two
  equal and opposite dislocations with Burgers vectors ${\bf b}$ and
  $-{\bf b}$, carries a unit of atom number, as can be seen by the
  deficiency of a single atom in the middle of the configuration. As
  the dislocations proliferate, the condensation of pairs of opposite
  dislocations $\left({\bf b},-{\bf b} \right)$ must always accompany
  the condensation of vacancies/interstitials, which correspond to
  terms of $\mathcal{L}_{\text{sf-dis}}$, the coupling between
  dislocation climb operators and vacancy or interstitial condensate
  (see Appendix A).}
\label{fig:sf-dis. coupling}
\end{figure}

As discussed in the Introduction, in the Mott-insulating commensurate
crystal phase, the dipole ${\bf p}_k$ can only move in the direction
perpendicular to ${\pv}_k$ while the along-dipole climb is forbidden
due to the U(1) particle-number conservation symmetry
(``glide-constraint''). Thus, we have only the glide motion,
$\Pi_{ij}^{\perp {\pv_k}} D_j \psi_{\pv_k}$, with,
${\bf D}=\grad +i p_k {\bf A}_k$, the covariant spatial derivative,
and
$\Pi_{ij}^{\perp {\pv_k}}=\delta_{ij}-\frac{p_{i,k} p_{j,k}}{p_k^2}$,
the transverse projection operator. However, the crystal also exhibits
scalar non-topological point defects, corresponding to deficiency and
excess in atom density, which permits the climb process of the dipoles
(dislocations), as shown in Fig. \ref{fig: Dislocation
  climb}. Combined with the bosonic statistics of the underlying
particles, the quantum crystal can first develop into a super-solid
phase (``incommensurate'' crystal), featuring both the crystalline and
the superfluid orders. The condensation of vacancies or interstitials
in the super-solid phase (as illustrated in
Fig.\ref{fig:sf-dis. coupling} is a bound state of opposite charge
dislocatons - a quadrupole), therefore, frees these symmetry-forbidden
climb
constraints\cite{MarchettiRadzihovsky,PretkoLRsymmetryEnrichedPRL2018,
  PretkoZhaiLRdualityPRB, Kumar19},
$\Pi_{ij}^{\parallel {\pv_k}} D_j \psi_{\pv_k}$, where
$\Pi_{ij}^{\parallel {\pv_k}}=\frac{p_{i,k} p_{j,k}}{p_k^2}$ is the
longitudinal projection operator. In Appendix A, we show in detail how
the dislocation-superfluid coupling alleviates the glide constraint
and converts the dislocations (dipoles) from subdimensional
quasi-particles to ordinary mobile defects, acted upon by the full
spatial derivative, ${\bf D}\psi_{\pv_k}$.

Introducing defect's core energy, $E_{j_{\mu}^b}$, to account for
lattice-scale physics, and the dipole (dislocation) charge density and
current density on a discrete lattice is given as a sum of their
discrete charges
\begin{eqnarray}
n^b_k({\bf r})
&=&\sum_{{\bf r}_n} d \cdot n^b_k({{\bf r}_n})\delta^2({\bf r}-{\bf r}_n),\\
{\bf j}^b
_k({\bf r}) &=& \sum_{{\bf r}_n} d \cdot {\bf j}^b_k ({{\bf r}_n})\delta^2({\bf r}-{\bf r}_n),
\end{eqnarray}
where $d$ is the lattice spacing for a 2D crystal, which is also the elementary charge of the gauge dipoles (unit of the dislocation charge), i.e., $|\pv_k|=p=d$.
The partition function is then given by
 \begin{equation}
 \begin{aligned}
      Z&=\sum_k\int \prod_{{\bf r}_n} dA_{\mu,k}({\bf r}_n) \sum_{j_{\mu,k}^b} \delta\left(\Delta_{\mu} j_{\mu,k}^b\right) \cdot e^{-S_{\text{M}}^{\text{cr}}}\\& \cdot e^{-\int dt \sum_{{\bf r}_n}\sum_k \left[\right.\frac{m_p}{2} \left( \partial_t \varphi_k+p A_{0,k}\right)^2 -d^2E_{j_{\mu,k}^b} |j_{\mu,k}^b({\bf r}_n)|^2+dA_{\mu,k} j_{\mu,k}^b \left.\right]}\\&\equiv \sum_k \int \prod_{{\bf r}_n} dA_{\mu,k}({\bf r}_n) d\varphi_k (\rv_n) \sum_{j_{\mu,k}^b} e^{- \tilde{S}_{\text{cr}}[A_{\mu,k}, j_{\mu,k}^b]}
 \end{aligned}
 \end{equation}
 with the action
 \begin{equation}
 \begin{split}
    \tilde{S}_{\text{cr}}=&\int dt \sum_{{\bf r}_n} \sum_k \bigg[\frac{m_p}{2} \left( \partial_t \varphi_k+p A_{0,k}\right)^2 -\tilde{E}_{j_{\mu,k}^b} |j_{\mu,k}^b({\bf r}_n)|^2\\&+d \left(A_{\mu,k}+\frac{1}{d}\Delta_{\mu} \varphi_{k} \right) j_{\mu,k}^b({\bf r}_n)\bigg]+S_{\text{M}}^{\text{cr}},
\end{split}
\end{equation}
where, $m_p$ is the effective inertia mass of the dipoles, $\tilde{E}_{j_{\mu,k}^b}=d^2 E_{j_{\mu,k}^b}$ in the discrete lattice, and, $\Delta_{\mu} j_{\mu,k}^b=j^b_{\mu,k}({\bf r}+{\bf \mu})-j^b_{\mu,k}({\bf r})$ and  $\Delta_{\mu}\varphi_k=\varphi_k ({\bf r}+{\bf \mu})-\varphi_k({\bf r})$, $\mu=x,y$, are the discrete lattice derivatives. Note the continuity equation $\Delta_{\mu} j_{\mu,k}^b=0$ is automatically satisfied when we integrate out $\varphi_k$, which is the phase of the $k-$flavor dipole field $\psi_k= |\psi_k| e^{i\varphi_k}$.
 After tracing over the 3-currents $j_{\mu,k}^b=(n^b_k, {\bf j}^b_k)$, we obtain
  \begin{equation}
 \begin{split}
    \tilde{S}_{\text{cr}}=&\int dt  \sum_{{\bf r}_n} \sum_k \bigg[\frac{m_p}{2} \left( \partial_t \varphi_k+p A_{0,k}\right)^2 \\&-g^b_k \cos\left(\Delta_{\mu} \varphi_k+p A_{\mu,k}\right)\bigg]+S_{\text{M}}^{\text{cr}},
\end{split}
\end{equation}
where $g^b_k=2 e^{-\tilde{E}_{j^b_k}}$, and we have approximated the resulting Villain potential by its lowest harmonic.

In the continuum limit, we have
\begin{equation}
    \tilde{S}_{\text{cr}}=\int dt \int d^2r \tilde{\mathcal{L}}_{\text{cr}},
\end{equation}
with the Lagrangian density $\tilde{\mathcal{L}}_{\text{cr}}$ given by
\begin{equation}
\begin{aligned}
    \tilde{\mathcal{L}}_{\text{cr}}=&\sum_k \left[\frac{\rho_k}{2} \left( \partial_t \varphi_k+p A_{0,k}\right)^2 -\tilde{g}^b_k \cos\left(\grad \varphi_k+p{\bf A}_k\right)\right]\\&+\mathcal{L}_{\text{M}}^{\text{cr}},
\label{Higgs crystal}
\end{aligned}
\end{equation}
where, $\tilde{g}^b_k=g^b_k/d^2=\frac{2}{d^2}e^{-\tilde{E}_{j^b_k}}$.

We then turn to an equivalent ``soft-spin'' description by noting that
this ordered phase action emerges from a corresponding quantum
Ginzburg-Landau theory for the complex order parameter, $\psi_k=
|\psi_k| e^{i\varphi_k}$, and write $\tilde{\mathcal{L}}_{\text{cr}}$
as,
\begin{equation}
    \tilde{\mathcal{L}}_{\text{cr}}=\sum_k\frac{1}{2} |\left(\partial_{\mu}+ipA_{\mu,k} \right)\psi_k|^2-V\left(\{|\psi_k|\}\right)+\mathcal{L}_{\text{M}}^{\text{cr}},
\label{soft-spin crystal}
\end{equation}
where, $\psi_k$ correspond to $\hat{\bf x}$- and
$\hat{\bf y}$-oriented dipole fields ($\hat{\bf y}-$ and
$\hat{\bf x}-$dislocations) in a square lattice, and
$V\left(\{|\psi_k|\}\right)$, is the Landau U(1)-invariant potential
for the quantum crystal, the form of which controls the type and
subsequence of phase transitions (See Section \ref{sec:Higgs
  transitions}).  It is straightforward to verify that
Eq.(\ref{soft-spin crystal}) reduces to Eq.(\ref{Higgs crystal}) when
the gapped Higgs-like magnitude degrees of freedom, $|\psi_k|$, whose
fluctuations are controlled by $V\left(\{|\psi_k|\}\right)$, are
integrated out, and therefore these two Lagrangians are equivalent.

\subsubsection{Quantum smectic}

Similar analysis applies also for the quantum smectic. The condensation of vacancies or interstitials in the super-smectic (``incommensurate" smectic) phase also endows the full mobility of dipoles (dislocations) in the smectic. 

Then, starting with $\tilde{\mathcal{L}}_{\text{sm}}$ given by
Eq.(\ref{Dual smectic}) and elevating dislocation and disclination
defects into dynamical charges, we add corresponding kinetic energies,
$\frac{\rho}{2} \left(\partial_t \varphi_x+pA_{0}\right)^2$, and
follow the same procedure as what have been done for the quantum
crystal above, which leads to the effective Lagrangian density of the
super-smectic in the continuum given by,
\begin{equation}
\begin{aligned}
    \tilde{\mathcal{L}}_{\text{sm}}=&\frac{\rho}{2} \left( \partial_t \varphi_x+p A_0\right)^2 -\tilde{g}_b \cos\left(\grad \varphi_x+p {\bf A}\right)+\mathcal{L}_{\text{M}}^{\text{sm}}\\=&\frac{1}{2} |\left(\partial_{\mu}+ipA_{\mu} \right)\psi_x|^2-V\left(|\psi_x|\right)+\mathcal{L}_{\text{M}}^{\text{sm}},
\label{Dual Super-smectic 1}
\end{aligned}
\end{equation}
where for concreteness we have taken the layers to be along the
$\hat{\bf x}$ axis, and replaced $\tilde{g}^b_x$ simply by
$\tilde{g}_b$ for the $\hat{\bf x}$-dipole ($\hat{\bf y}$-dislocation)
fugacity, and $A_{\mu,x}$ simply by $A_{\mu}$. In the second form, we
have written it as an equivalent ``soft-spin'' description in terms of
$\psi_x= |\psi_x| e^{i\varphi_x}$, with the Landau U(1)-invariant
potential, $V\left(|\psi_x|\right)$, which is equivalent to the first
form when the gapped Higgs magnitude degree of freedom,
$|\psi_x|$, whose fluctuations are controlled by
$V\left(|\psi_x|\right)$, is integrated out.

\subsection{Crystal-to-smectic and smectic-to-nematic transitions}
\label{sec:Higgs transitions}
The quantum crystal phase can go through a fully isotropic melting
transition, mediated by dislocations, into a hexatic (or nematic)
phase, or through a multi-stage anisotropic transition, first
partially melting into a smectic phase, depending on the form of the
Landau potential $V(\{|\psi_k|\})$. A simple discussion based on
Ginzburg-Landau theory of continuous phase transitions is given in the
following.

Considering a 2D square lattice, the U(1)-invariant Landau potential $V (\{|\psi_k|\})$, satisfying the symmetries of the system, expanded to fourth-order of the dipole fields $\psi_k$, is given by 
 \begin{equation}
 \begin{split}
     V&=\frac{\alpha}{2} \sum_{k} |\psi_k|^2 +\frac{\beta}{4} \sum_{k}|\psi_{k}|^4 +\frac{\beta'}{2} |\psi_x|^2 |\psi_y|^2\\
     &\equiv \frac{\alpha}{2} |\Psi|^2+\frac{\beta}{4} |\Psi|^4+\frac{1}{2} \left( \beta'-\beta\right)|\psi_x|^2|\psi_y|^2,
\end{split}
 \end{equation}
 where, $\beta, \beta' >0$, and we have defined the vector complex
 order parameter $\Psi=(\psi_x,\psi_y)$. Note that by these two forms,
 we can regard this potential as two identical complex Ising models
 coupled together, or a complex XY model with the originally
 rotational symmetry broken down to a rectangular one. As usual, at
 mean-field level the phase transition takes place at $\alpha=0$, as
 $\alpha$ changes its sign.
 
\subsubsection{Crystal-to-nematic transition}
 
We first consider the case of $\beta >\beta'$, when it is
energetically favorable for both flavors of dipole fields to condense
with the same expectation value
$|\psi_x|=|\psi_y|=\sqrt{\frac{|\alpha|}{\beta+\beta'}}$,
corresponding to the nematic phase. This crystal-to-nematic
transition, analogue of the crystal-to-hexatic melting in a 2D
classical crystal, has been discussed in
Ref. \onlinecite{PretkoZhaiLRdualityPRB}, formulated as a
Ginzburg-Landau theory of tensor superconductors, i.e., dipole fields
coupled to the symmetric tensor gauge field. However, as pointed out
in the Introduction, that tensor-only gauge theory formulation of the
Mott-insulating ``commensurate'' crystal, fails to capture the full
dipole mobility endowed by the condensation of vacancies and
interstitials. Here, we take a complementary coupled vector gauge
theory description\cite{RadzihovskyHermeleVectorGaugePRL2020} and
condense both $\hat{\bf x}$- and $\hat{\bf y}$-dipoles,
$\psi_x \neq 0, \psi_y \neq 0$. Via Anderson-Higgs mechanism this gaps
out all gauge field flavors, $A_{\mu, k}, k=x,y$. These can then be
safely integrated out at low energies at wavelengths longer than
$\sqrt{\frac{2m_p}{\rho_k p^2}}$ (i.e., the penetration length of
$A_{\mu,k}$), thereby reducing the Maxwell Lagrangian of the crystal
to that of a form described by the rotational gauge field $a_{\mu}$
only. With details relegated to Appendix B, the result is given by,
 \begin{equation}
 \begin{split}
     \mathcal{L}_{\text{M}}^{\text{cr}} \left(A_{\mu, k}=0, a_{\mu} \right) \approx & \mathcal{L}_{\text{M}}^{\text{nm}} \left(a_{\mu} \right)\\=&\frac{1}{2}K^{-1}\left(\partial_t \av-\grad a_0\right)^2-\frac{1}{2} \left(\grad \times \av\right)^2,
\end{split}
\end{equation}
which corresponds to setting, $A_{\mu, k} \approx 0$, to the lowest
order, and neglecting anisotropies of the resulting nematic state.
 
As demonstrated by foundational papers on duality \cite{dasgupta,
  fisher}, such Abelian gauge theory is dual to a quantum XY model of
the nematic given by,
\begin{equation}
     \mathcal{L}_{\text{nm}}=\frac{1}{2} \left(\partial_t \theta \right)^2-\frac{1}{2} K \left(\grad \theta \right)^2,
\end{equation}
where $\theta$ is the layer orientation, confirming a consistency with
our physical expectations.
 
\subsubsection{Crystal-to-smectic and smectic-to-nematic transitions}
 
We next consider the case of $\beta <\beta'$, when it is favorable to
have only one flavor of dipole fields condensed, with $\psi_x=0$ and
$|\psi_y|=\sqrt{\frac{|\alpha|}{\beta}}$, or, $\psi_y=0$ and
$|\psi_x|=\sqrt{\frac{|\alpha|}{\beta}}$, corresponding to the
crystal-to-smectic phase transition, which restores translational
symmetry in only one direction. With our interest in the quantum
smectic, in the following we will thus focus on the $\beta<\beta'$
case, and dualize the quantum smectic alternatively, via
Anderson-Higgs mechanism, gapping out one of the flavors of the gauge
fields in this smectic phase, formulated in a ``soft-spin''
description of a 2D quantum crystal.

The crystal-smectic partial melting transition corresponds to
condensation of one flavor of the dipole fields, that according to
Fig. \ref{fig:phase transition}, we take to be $\hat{\bf y}$-dipoles,
i.e. $\psi_y \neq 0$. Within this Higgs phase, the corresponding $y$
flavor gauge field $A_{\mu, y}$ is gapped out, and can be safely
integrated out in the low-energy regime with wavelengths of
excitations much greater than $\sqrt{\frac{2m_p}{\rho_y p^2}}$ (i.e.,
the penetration length of $A_{\mu,y}$ ). To lowest order, it
corresponds to $A_{\mu, y} \approx 0$, reducing the crystal's Maxwell
Lagrangian to that of a smectic,
\begin{equation}
    \mathcal{L}_{\text{M}}^{\text{cr}} \left(A_{ \mu, x}, A_{\mu, y}=0, a_{\mu} \right) \approx \mathcal{L}_{\text{M}}^{\text{sm}} \left(A_{\mu, x}, a_{\mu} \right).
\end{equation}
The detailed derivations are given in Appendix B. Subsequent melting with a condensation of $\hat{\bf x}-$dipoles ($\hat{\bf y}-$dislocations) leads to a single vector gauge theory for $a_{\mu}$, with both $A_{\mu,k}$ gapped out, in the low-energy regime with wavelengths of excitations much greater than $\sqrt{\frac{2m_p}{\rho_x p^2}}$, i.e., 
\begin{equation}
\begin{split}
    \mathcal{L}_{\text{M}}^{\text{sm}} \left(A_{ \mu, x}=0, a_{\mu} \right) \approx & \mathcal{L}_{\text{M}}^{\text{nm}} \left(a_{\mu} \right)\\=&\frac{1}{2}K^{-1}\left(\partial_t \av-\grad a_0\right)^2-\frac{1}{2} \left(\grad \times \av\right)^2,
\end{split}
\end{equation}
to the lowest order. As expected, this corresponds to the dual of the
quantum XY model of the nematic state.
 
\subsection{Vector to tensor gauge theory redux in low-energy limit}
The Maxwell part of a crystal, given by Eq.(\ref{Maxwellcrystal}), is
gauge invariant under the following transformation,
\begin{equation}
    \begin{split}
        \left(A_{0k}, {\bf A}_k \right) &\to \left( A_{0k}+\partial_t \chi_k, {\bf A}_k+\grad \chi_k\right),\\
        \left(a_0,a_k\right) &\to \left(a_0+\partial_t \phi, a_k+\partial_k \phi+\chi_k \right).
    \end{split}
\end{equation}
As demonstrated explicitly in Ref. \onlinecite{RadzihovskyHermeleVectorGaugePRL2020}, the enlarged gauge redundancy
allows us to completely eliminate ${\bf a}$ by choosing $\chi_k=a_k$,
as a result of which, the term $\frac{1}{2}\left(\grad \times {\bf
    a}-\hat{\bf z}\times {\bf A}_k\right)^2$ reduces to $\frac{1}{2}
\left(\epsilon_{ik}A_{ik}\right)^2$, thereby gapping out the
antisymmetric component $\epsilon_{ik}A_{ik}$ at energies well below
this gap, i.e., with length scales greater than
$\sqrt{\chi}$. Furthermore, the electric field term $\frac{1}{2}K^{-1}
(\partial_t {\bf a}-\grad a_0- A_{0k} \hat{\bf e}_k)^2$ reduces to
$\frac{1}{2}K^{-1} (\grad a_0+ A_{0k} \hat{\bf e}_k)^2$ under this
transformation, enforcing $A_{0k}=-\partial_ka_0$ at low enough
energies with length scales greater than
$\sqrt{\frac{K}{\chi}}$. Therefore, the dual coupled U(1) vector
gauge theory for a quantum crystal, reduces to the dual tensor gauge
theory in the low-energy limit, with,
$\mathcal{L}_{\text{M}}^{\text{cr}}$, reduces to that in the tensor
gauge theory, described by,
\begin{equation}
\begin{aligned}
    \mathcal{L}_{\text M}^{\text{cr}}=&\frac{1}{2}\chi^{-1}
    \left(\partial_t A_{ik}+\partial_i\partial_k a_0
    \right)^2-\frac{1}{2} \left(\epsilon_{ji}\partial_jA_{ik}
    \right)^2 \\
    =&E_{ik} \left( \partial_t A_{ik}+\partial_i\partial_k a_0\right)- \frac{1}{2}\chi E_{ik}^2-\frac{1}{2} B_k B^k,
\label{L_cr low energies}
\end{aligned}
\end{equation}
where $A_{ik}$, is a  rank-2 symmetric tensor field, which corresponds to the $i^{\text{th}}$ component of $k$-flavor vector gauge field ${\bf A}_k$ in the coupled vector gauge theory, 
$a_0$ is a scalar field with $\partial_k a_0$ corresponding to $A_{0k}$ in the vector gauge theory, $B_k=\epsilon_{ji}\partial_jA_{ik}$, and $E_{ik}$ is the electric tensor field canonically conjugate to $A_{ik}$. 

In contrast, within the smectic Higgs phase, corresponding to the condensation of $\hat{\bf y}-$dipoles ($\hat{\bf x}-$dislocations), we cannot eliminate ${\bf a}$ completely, since we have already made a gauge choice with $\chi_y=\frac{1}{p}\varphi_y$,
\begin{equation}
     A_{iy} \to A_{iy}+\frac{1}{p} \partial_i \varphi_y,\; A_{0y} \to A_{0y}+\frac{1}{p} \partial_t \varphi_y,\; a_y \to a_y+\frac{1}{p}\varphi_y,
 \end{equation}
 to absorb the phase $\varphi_y$ of the condensed dipole field
 $\psi_y=\sqrt{\rho_y} e^{i\varphi_y}$ into the gauge fields $A_{\mu,
   y}$.  Integrating out the gapped gauge field components, $A_{iy}$
 and $\partial_y a_0$, reduces $\mathcal{L}_{\text{M}}^{\text{cr}}$,
 in a condensate of $\hat{\bf y}-$dipoles, to that of a smectic,
\begin{equation}
\begin{aligned}
    \mathcal{L}_{\text M}^{\text{cr}}=&\frac{1}{2}\chi^{-1} \left(\partial_t A_{ix}+\partial_i\partial_x a_0 \right)^2-\frac{1}{2} \left(\epsilon_{ji}\partial_jA_{ix} \right)^2\\&+\frac{1}{2} K^{-1} \left(\partial_t {\bf a}-\grad a_0-A_{0x} \hat{\bf x} \right)^2+\frac{1}{2} \left(\epsilon_{ij}\partial_i a_j+A_{yx}\right)^2 \\=& -{\bf E}_x \cdot \left( \partial_t {\bf A}_x-\grad A_{0x}\right)-\frac{1}{2} \chi {\bf E}_x^2-\frac{1}{2}\left(\grad \times {\bf A}_x \right)^2\\&- {\bf e} \cdot \left(\partial_t {\bf a}-\grad a_0-A_{0x} \hat{\bf x}\right)-\frac{1}{2} K {\bf e}^2\\&-\frac{1}{2}\left(\grad \times {\bf a}+\hat{\bf x}\times {\bf A}_x\right)^2,
\end{aligned}
\end{equation}
which matches exactly with Eq.(\ref{Maxwellsmectic}) after setting $A_{0x}=-\partial_x a_0$ and dropping the `x' index.

\begin{figure}[htbp]
\centering
 \hspace{0in}\includegraphics*[width=0.5\textwidth]{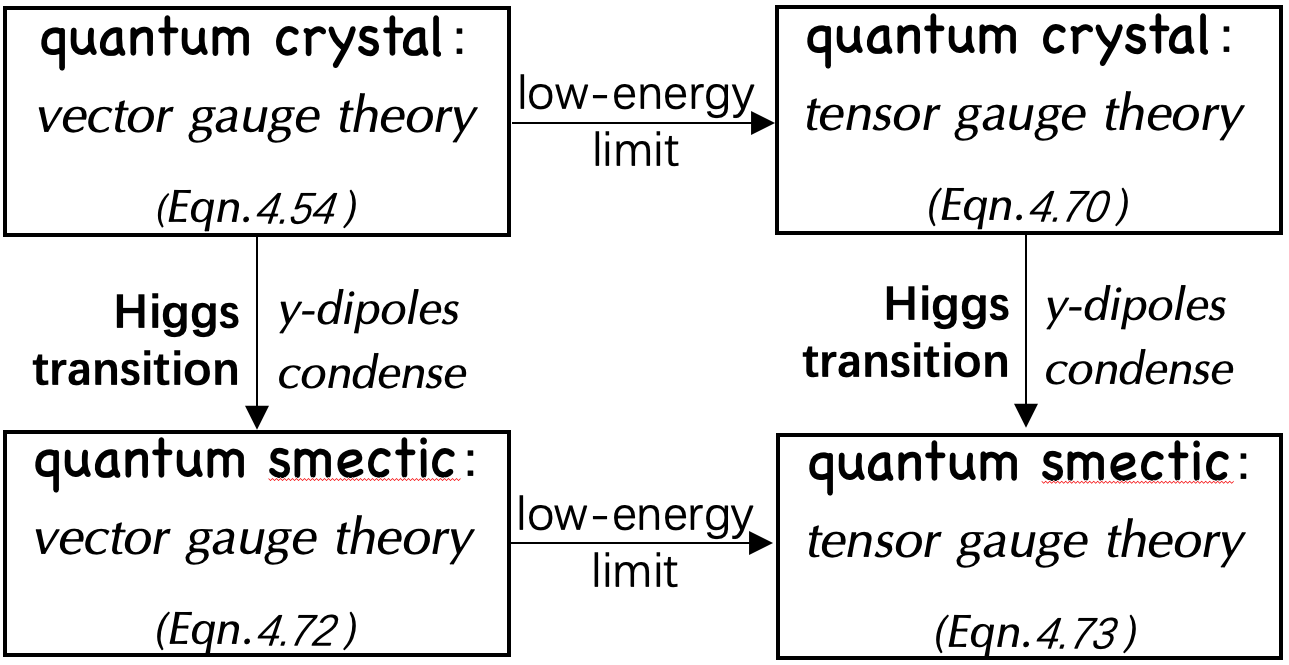}
 \caption{Vector gauge theory of a 2D quantum crystal and smectic, in
   the low-energy limit reduce to their corresponding tensor gauge
   theory forms.}
\label{fig:vectortotensor}
\end{figure}

We may also explore a 2D quantum smectic at low energies by similar
analysis, as what has been done for a crystal in Eq.(\ref{L_cr low
  energies}). In the smectic case, choosing $\chi=a_x$ in the gauge
transformation (\ref{gauge transform}) allows us to eliminate $a_x$
completely, as a result of which, the term
$\frac{1}{2}\left(\grad \times {\bf a}+\hat{\bf x}\times {\bf
    A}\right)^2$ of $\mathcal{L}_{\text{M}}^{\text{sm}}$ given by
(\ref{Maxwellsmectic}), reduces to,
$\frac{1}{2}\left(\partial_x a_y+A_y \right)^2$, thereby enforcing
$A_y=-\partial_x a_y$ at sufficiently low energies, with length scales
greater than $\sqrt{\chi}$. Furthermore, the term
$\frac{1}{2} K^{-1} \left(\partial_t {\bf a}-\grad a_0-A_0 \hat{\bf
    x}\right)^2$, reduces to
$\frac{1}{2} K^{-1} \left[\left(\partial_t a_y-\partial_y
    a_0\right)^2+\left(\partial_x a_0+A_0 \right)^2 \right]$ under
this transformation, enforcing $A_0=-\partial_x a_0$ in low energy
regime with length scales greater than
$\sqrt{\frac{K}{\chi}}$. Therefore, in the low-energy limit,
$\mathcal{L}_{\text{M}}^{\text{sm}}$ reduces to,
\begin{equation}
\begin{split}
    \mathcal{L}_{\text{M}}^{\text{sm}}=&\frac{1}{2} \chi^{-1} \left( \partial_t A_i+\partial_i\partial_x a_0\right)^2-\frac{1}{2}\left(\grad \times {\bf A} \right)^2\\&+\frac{1}{2} K^{-1} \left(\partial_t a_y-\partial_y a_0\right)^2\\=&-E_{ix} \left(\partial_t A_i+\partial_x\partial_x a_0\right)-\frac{1}{2}\chi E_{ix}^2-\frac{1}{2}B_x^2\\&+\frac{1}{2} K^{-1} \left(\partial_t a_y-\partial_y a_0\right)^2,
\end{split}
\end{equation}
where $E_{ix}$ is a tensor gauge field, corresponding to the
$i^{\text{th}}$ component of the vector field ${\bf E}_x$.  This
vector- to tensor-gauge theory reduction at low energies is
illustrated in Fig.\ref{fig:vectortotensor}.

\section{Classical limit of smectic-gauge theory duality}
\label{sec:Classical limit}
\subsection{2D classical smectic duality}
  As a consistency check on our quantum dual theory for a 2D quantum smectic, we anticipate that the classical smectic theory must emerge as the classical limit of the above duality, as we demonstrate explicitly below. 
  
  The elasticity of a 2D classical smectic is given by the Hamiltonian density,
  \begin{equation}
  \begin{split}
      \mathcal{H}_{\text{sm}}&=\frac{1}{2} \chi \left(\grad u-\theta \hat{\bf x}\right)^2+\frac{1}{2}K \left(\grad \theta \right)^2\\&=\frac{1}{2} \chi^{-1} {\bf \sigma}^2-i {\bf \sigma} \cdot \left(\grad u- \theta \hat{\bf x} \right)+\frac{1}{2} K^{-1} \jv^2-i  \jv \cdot \grad \theta.
  \end{split}
  \end{equation}
  where we have introduced two Hubbard-Stratonovich fields ${\bf \sigma}$ and $\jv$ to decouple the two elastic terms.
  
 We derive its classical dual by integrating out the smooth part of $u$ and $\theta$, which leads to the constraints:
  \begin{equation}
      \begin{split}
          \partial_i \sigma_i&=0,\\
          \partial_i j_i+ \sigma_x&=0.
      \end{split}
  \end{equation}
  The first equation can be solved in terms of a scalar potential $\phi$, $\sigma_i=\epsilon_{ij}\partial_j \phi$, which inside the second constraint gives,
  \begin{equation}
      \partial_i (j_i+\epsilon_{xi} \phi)=0,
  \end{equation}
  that is then solved by introducing another potential $\alpha$, $j_i=\epsilon_{ij}\partial_j \alpha-\epsilon_{xi} \phi$.
  
  Substituting $\sigma_i$ and $j_i$ back into the original Hamiltonian, and integrating by parts, lead to
  \begin{equation}
      \begin{split}
         \tilde{\mathcal{H}}_{\text{sm}}=&\frac{1}{2} \chi^{-1} \left(\grad \phi\right)^2-i\epsilon_{ij} \partial_j \phi \left(\partial_i u-\delta_{ix} \theta \right)\\&+\frac{1}{2} K^{-1} \left( \epsilon_{ij}\partial_j \alpha-\epsilon_{xi}\phi \right)^2-i \left( \epsilon_{ij}\partial_j \alpha-\epsilon_{xi}\phi \right) \partial_i \theta\\=&\frac{1}{2} \chi^{-1} \left(\grad \phi\right)^2+i \phi \epsilon_{ij}\partial_j \left(\partial_i u-\delta_{ix} \theta \right)\\&+\frac{1}{2}K^{-1} \left(\grad \alpha+\phi \hat{\bf x} \right)^2+i\theta \epsilon_{ij}\partial_i \partial_j \alpha+i \phi \epsilon_{xi}\partial_i \theta\\=&\frac{1}{2} \chi^{-1} \left(\grad \phi\right)^2+\frac{1}{2}K^{-1} \left(\grad \alpha+\phi \hat{\bf x} \right)^2-i \phi b-i\alpha s,
      \end{split}
  \end{equation}
  where we have defined $b=\epsilon_{ij}\partial_i\partial_j u$ and $s=\epsilon_{ij}\partial_i\partial_j \theta$, as the dislocation and disclination densities respectively.
  
Low-energy regime, $\frac{1}{2}K^{-1} \left(\grad \alpha+\phi \hat{\bf x} \right)^2 \approx k_BT $, i.e., at length scales greater than $\sqrt{Kk_BT}$, integrating over $\phi$ , to lowest-order, sets $\phi=\partial_x\alpha$, and therefore, gives
  \begin{equation}
      \tilde{\mathcal{H}}_{\text{sm}}=\frac{1}{2}\chi^{-1} \left(\partial_x^2 \alpha \right)^2+\frac{1}{2} K^{-1} \left(\partial_y \alpha\right)^2-i \partial_x \alpha b-i \alpha s,
 \label{Classical Dual}
  \end{equation}
  which, as expected, turns out to be the electrostatic limit of the
  quantum smectic duality, i.e., Eq.(\ref{Dual smectic}) and
  Eq. (\ref{Maxwellsmectic}), with $\phi=A_0$ and $\alpha=a_0$.
  
  Focusing on dislocations and neglecting the high energy disclination defects, we can straightforwardly integrate out $\alpha({\bf r})$ in the partition function, obtaining  a dislocation Coulomb gas Hamiltonian
  \begin{equation}
      H_{\text{b}}=\frac{1}{2}\int \frac{d^2q}{(2\pi)^2} b({\bf q}) \tilde{K}({\bf q}) b(-{\bf q}),
  \end{equation}
  with,
  \begin{equation}
      \tilde{K}({\bf q})=\frac{d^2}{a^2}\frac{K q_x^2}{q_y^2+\lambda^2 q_x^4}+2 E_b,
     \label{2D core}
  \end{equation}
  where, $E_{\text{b}}$ is the defect core energy, d is the layer spacing, and a is the lattice spacing between atoms within the layers, and ``penetration" length $\lambda$ is defined as, $\lambda^2 \equiv K/\chi$.
  Thus, the dislocations Coulomb gas Hamiltonian in real space reduces to,
  \begin{equation}
      H_{\text{b}}=\frac{1}{2}\int_{{\bf r}_1,{\bf r}_2} U({\bf r}_1-{\bf r}_2) b({\bf r}_1) b({\bf r}_2)+ E_{\text{b}} \int_{\bf r} b^2 ({\bf r}),
  \label{dislocation Coulomb gas}
  \end{equation}
  where,
  \begin{equation}
      U({\bf r})=\frac{1}{4} \frac{d^2}{a^2} \chi \left( \frac{\lambda}{\pi |y|} \right)^{1/2} e^{-x^2/4\lambda |y|},
  \end{equation}
  as first found by Toner and Nelson in Ref. \onlinecite{TonerNelson}. The complete Hamiltonian contains also a smooth phonon part, $H_0$, depending only the smooth, single-valued part, $\tilde{u}$, of the displacement, $u=\tilde{u}+u^s$, i.e.,
\begin{equation}
      H=H_0+H_b.
\end{equation}
  with $H_0$ given by
\begin{equation}
  H_0=\frac{1}{2}\int d^2r \left[ \chi \left(\partial_y \tilde{u} \right)^2+K \left(\partial_x^2 \tilde{u} \right)^2\right]
\end{equation}
With this Hamiltonian, we can study the effects of phonons and
dislocations at finite temperatures on translational and orientational
orders.
  
Effect of phonon fluctuations on the translation order is expressed in
terms of correlations in the order parameter, $\psi\left({\bf
    r} \right) = |\psi| e^{-i q_0 u({\bf r})}$ as,
   \begin{equation}
\begin{split}
\langle \psi ({\bf r}) \psi^*({\bf 0})\rangle &\sim \langle e^{-iq_0 \left[u(\rv)-u({\bf 0}) \right]} \rangle\\&=e^{-\frac{1}{2}q_0^2 \langle [u(\rv)-u({\bf 0})]^2\rangle}
\\&\sim \left\{\begin{array}{ll}
 \exp \left(-\frac{q_0^2k_BT}{\chi} \sqrt{\frac{|y|}{4\pi \lambda}}\right), \mbox{ for $ x^2 \ll \lambda y$},
\\ \exp \left(-\frac{q_0^2k_BT}{4\chi\lambda} |x|\right), \mbox{ for $ x^2 \gg \lambda y$,}
 \end{array}\right.
\end{split}
\end{equation}
where we have used the fact that 
\begin{equation}
\begin{split}
  \langle[u(\rv)-u({\bf 0})]^2\rangle&=\int_{\qv}\int_{\qv'} \left(e^{i \qv \cdot \rv}-1 \right) \left(e^{i \qv' \cdot \rv}-1 \right)\langle u(\qv) u(\qv') \rangle
  \\&=\int_{\qv} \left(2-2\cos (\qv \cdot \rv)\right) \frac{k_BT}{\chi \left(q_y^2+\lambda^2 q_x^4\right)}\\&\sim
  \left\{\begin{array}{ll}
 \frac{k_BT}{\chi} \sqrt{\frac{|y|}{\pi \lambda}}, \mbox{ for $ x^2 \ll \lambda y$},
\\ \frac{k_BT}{2\chi\lambda} |x|, \mbox{ for $ x^2 \gg \lambda y$.}
 \end{array}\right.
\end{split}
\end{equation}
 Smectic layers orientational order is expressed in terms of correlations in the nematic-like order parameter, ${\bf N} ({\bf r})= \left(\cos \theta ({\bf r}), \sin \theta({\bf r}) \right)$, as,
 \begin{equation}
 \begin{split}
    \lim_{r\to \infty} \langle {\bf N} ({\bf r}) \cdot {\bf N}({\bf 0}) \rangle &=\lim_{r\to \infty}\langle \cos\left[\theta (\rv)-\theta({\bf 0}) \right]\rangle\\&=\lim_{r\to \infty} e^{-\frac{1}{2} \langle\left[\theta (\rv)-\theta({\bf 0}) \right]^2 \rangle}\\&= e^{-\frac{K_BT\Lambda}{\chi \lambda a}}=\text{constant},
 \end{split}
 \end{equation}
 where we have used 
 \begin{equation}
 \begin{split}
    &\langle\left[\theta (\rv)-\theta({\bf 0}) \right]^2 \rangle = \langle\left[\partial_x u (\rv)-\partial_x u({\bf 0}) \right]^2 \rangle\\&=-\int_{\qv}\int_{\qv'} \left(e^{i \qv \cdot \rv}-1 \right) \left(e^{i \qv' \cdot \rv}-1 \right) q_x q_x'\langle u(\qv) u(\qv') \rangle\\&=\int_{\qv} \left(2-2\cos (\qv \cdot \rv)\right) \frac{q_x^2 k_BT}{\chi q_y^2+K q_x^4}\\&\sim\frac{2k_BT }{ \chi \lambda a}, \mbox{ for $r \to \infty$.}
\end{split}
\end{equation}
with $\Lambda=\frac{2\pi}{a}$, a convenient cutoff.  Therefore, in a
2D classical smectic at nonzero temperature, the translational order
is destroyed by thermal phonon fluctuations\cite{Landau37,
  Peierls36,TonerNelson}, while the orientational order persists even
in the presence of thermally excited phonons, destroyed only at higher
temperatures by proliferation of dislocations.
 
In presence of unbound dislocations, appearing at density, $n_d \sim
e^{-E_c/(k_{\text{B}}T)}$, in thermal equilibrium, the effective
elasticity in Debye-Huckel approximation, reduces to that of a nematic
at scales greater than, $\xi_{\text{D}} \sim e^{-E_c/(2k_{\text{B}}
  T)}$,
\begin{equation}
    F_{\text{N}}=\frac{1}{2} K(T) \int d^2 r \left(\grad \theta \right)^2,
\end{equation}
and the correlations in orientational order become decay algebraically,
\begin{equation}
     \lim_{r\to \infty} \langle {\bf N} ({\bf r}) \cdot {\bf N}({\bf 0}) \rangle  \sim  r^{-\eta (T)},
\end{equation}
with, $\eta(T)=2k_{\text{B}}T/[\pi K(T)]$. Therefore, a 2D smectic is
unstable to thermal fluctuations, driven into a nematic fluid at any
nonzero temperatures. At high temperature, the nematic to isotropic
liquid transition, driven by unbinding of disclinations, is described
by Kosterlitz and Thouless, with \cite{KT,Stein78},
\begin{equation}
    \eta \left(T_c^- \right)=\frac{1}{4},
\end{equation}
at the critical temperature $T_c$.

\begin{figure}[htbp]
  \centering
  \subfigure[]{
    \includegraphics[width=0.505\textwidth]{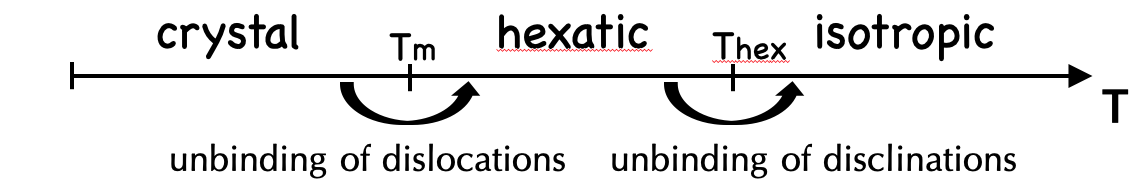}}
    \label{fig:crystalmelting}
    \hspace{0.0cm}
  \subfigure[]{
    \includegraphics[width=0.5\textwidth]{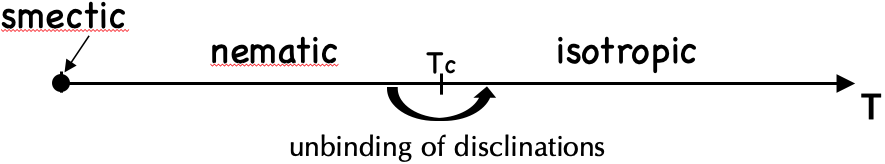}}
    \label{fig:smecticmelting}
    \caption{Different phases involved in the melting transitions of a
      triangular crystal and a smectic in two dimensions. (a) A
      triangular crystal at low temperatures, which first melts at
      $T_{\text{m}}$ via an unbinding of dislocations into a hexatic,
      and the sixfold orientational order in the hexatic phase is
      destroyed at a higher temperature $T_{\text{hex}}$ by 
      unbinding of disclinations. (b)  A zero-temperature
      smectic, that is unstable and driven into a nematic at any
      nonzero temperatures. The nematic phase transforms into an
      isotropic fluid via a further disclination unbinding transition
      at $T_c$. [Figures adapted from Reference
      \onlinecite{TonerNelson}.]}
\label{fig: Crystal+Smectic meltings}
\end{figure}

Motivated by our formulation of two-dimensional melting of a classical
crystal\cite{KT,Nelson79,Halperin79,HalperinOstlund,sineGordon}, via a
dual theory in terms of a higher derivative vector sine-Gordon model
\cite{Zhai19}, we expect to find the analogous description for the 2D
smectic. To this end, we express the dislocation and disclination
densities in terms of a sum of their discrete charges as,
\begin{eqnarray}
  \begin{split}
      b({\bf r})&=& \sum_{{\bf r}_n} b_{{\bf r}_n} \delta^2 ({\bf r}-{\bf r}_n),\\
      s({\bf r})&=&\sum_{{\bf r}_n} s_{{\bf r}_n} \delta^2 ({\bf r}-{\bf r}_n).
  \end{split}
\end{eqnarray}
In terms of these discrete topological defect charges, the Hamiltonian is given by
\begin{equation}
\begin{split}
    H_{\text{sm}}=&\frac{1}{2} \int_{\bf r} \left[\chi^{-1} (\partial_x^2 \alpha)^2+K^{-1} (\partial_y \alpha)^2 \right]\\&+\sum_{{\bf r}_n} \left[E_b b_{{\bf r}_n}^2+E_s {s_{{\bf r}_n}} \right]-\sum_{{\bf r}_n} \left[ i \partial_x \alpha b_{{\bf r}_n}+i 2\pi \alpha  s_{{\bf r}_n}\right].
\end{split}
\end{equation}
Following a standard analysis, summing over the charges, we obtain the
dual sine-Gordon Hamiltonian,
\begin{equation}
\begin{split}
    \tilde{H}_{\text{sm}}=&\int_{\bf r} \bigg[\frac{1}{2} \chi^{-1} (\partial_x^2 \alpha)^2+\frac{1}{2} K^{-1} (\partial_y \alpha)^2-g_b \cos (b  \partial_x \alpha)\\&-g_s \cos (2\pi \alpha) \bigg].
\end{split}
\label{2D sine-Gordon}
\end{equation}
where $g_b=\frac{2}{a^2} e^{-a^2 E_b}$ and
$g_s=\frac{2}{a^2} e^{-E_s}$, which provides a transparent description
of the continuous two-stage melting in terms of the
renormalization-group relevance of two cosine operators that control
the sequential unbinding of dislocations and disclinations,
respectively corresponding to the smectic-to-nematic and
nematic-to-isotropic fluid transitions.  The resulting phase diagram
is illustrated in Fig.\ref{fig: Crystal+Smectic meltings}.
  
 Because of the second-order
Laplacian elasticity, standard analysis around the Gaussian fixed line
$g_b = g_s = 0$ shows that, the mean-squared fluctuations of
$\alpha(\rv)$ is given by
\begin{equation}
    \langle[\alpha(\rv)-\alpha({\bf 0})]^2\rangle \sim
  \left\{\begin{array}{ll}
 \frac{k_BT}{\chi^{-1}} \sqrt{\frac{\lambda |y|}{\pi}}, \mbox{ for $ x^2 \ll \lambda^{-1} y$},
\\ \frac{k_BT}{2\chi^{-1}} \lambda |x|, \mbox{ for $ x^2 \gg \lambda^{-1} y$,}
 \end{array}\right.
\end{equation}
which leads to an exponentially (as opposed to power-law in a
conventional sine-Gordon model) vanishing Debye-Waller factor,  $\langle e^{i2\pi\alpha (\rv)} e^{-i 2\pi\alpha ({\bf 0})} \rangle=e^{-2\pi^2 \langle [\alpha(\rv)-\alpha({\bf 0})]^2\rangle}$, and in
turn to a strongly irrelevant disclination cosine, $g_s$, that can
therefore be neglected. In contrast, mean-squared fluctuations of $\partial_x \alpha(\rv)$ is,
\begin{equation}
    \langle\left[\partial_x \alpha (\rv)-\partial_x \alpha({\bf 0}) \right]^2 \rangle \sim \text{constant},
\end{equation}
for large $r$, and orientational correlation therefore given by,
\begin{equation}
\begin{aligned}
  \langle e^{ib\cdot \partial_x \alpha (\rv)} e^{-ib\cdot \partial_x \alpha ({\bf 0})} \rangle&=e^{-\frac{1}{2}b^2 \langle [\partial_x \alpha(\rv)- \partial_x \alpha({\bf 0})]^2\rangle}\\& \sim \text{constant}.
\end{aligned}
\end{equation}
This therefore leads to the conclusion that the dislocation cosine,
$g_b$, is always relevant.  At sufficiently long scales, dislocation
cosine in Eq.(\ref{2D sine-Gordon}) reduces to a harmonic potential
for $\partial_x \alpha$,
$-g_b\cos(b \partial_x\alpha)\simeq \oh g_b
b^2(\partial_x\alpha)^2$. The effective Hamiltonian is then given by
\begin{equation}
 \tilde H_{\text{nm}} \simeq\int_\rv \left[\oh
   K^{-1}(\partial_y\alpha)^2 + \oh
  g_b b^2(\partial_x \alpha)^2 - g_s\cos(2\pi\alpha)\right], 
\end{equation}
where we have neglected the $\chi^{-1}$ ``curvature'' elasticity
relative to the gradient one encoded in $g_b$, and restored the
disclination cosine operator $g_s\cos(2\pi \alpha)$. The resulting
conventional sine-Gordon model in $\alpha$ can then exhibit the
second KT-like ``roughening'' transition, capturing the
nematic-to-isotropic fluid transition, associated with the unbinding of
disclinations, with well-known standard KT phenomenology.

\subsection{3D classical smectic duality}
Motivated by the correspondence of a $(2+1)$D quantum smectic and a 3D
classical smectic, and the extensively studied 3D nematic to smectic-A
transition \cite{deGennes72, HalperinMa, Helfrich78, NelsonToner,
  Lubensky81, Grinstein86, Toner82}, we formulate a dual gauge theory
of a 3D classical smectic, akin to a mapping of a 3D classical XY
model onto a classical charged superconductor.\cite{dasgupta,fisher}

The elasticity of a 3D classical smectic with its layers along $xy$
plane, is captured by the Hamiltonian density,
 \begin{equation}
 \begin{split}
      \mathcal{H}_{\text{sm}}^{\text{3d}}&=\frac{1}{2}\chi \left(\grad u+ \delta {\bf n} \right)^2+\frac{1}{2}K \left(\grad \delta{\bf n} \right)^2\\&=\frac{1}{2\chi} {\bf \sigma}^2-i {\bf \sigma} \cdot \left(\grad u+ \delta {\bf n} \right)-\frac{1}{2K}  \jv \frac{1}{\grad^2} \jv-i  \jv \cdot \delta \nv ,
 \label{3D smectic H}
\end{split}
\end{equation}
 where $\delta \nv=\nv-\hat{\bf z}=\left(\delta n_x, \delta
   n_y,0\right)$ represents fluctuations in layer orientation, and we
 introduced two Hubbard-Stratonovich fields ${\bf \sigma}$ and $\jv$
 to decouple the two elastic terms. The Hamiltonian density in
 Eq. (\ref{3D smectic H}) is equivalent to the standard smectic form,
  \begin{equation}
      \mathcal{H}_{\text{sm}}^{\text{3d}}=\frac{1}{2}\chi \left(\nabla_z u \right)^2+\frac{1}{2}K \left(\grad_{\perp}^2 u \right)^2,
 \end{equation}
 in the low-energy limit, where the orientational degree of freedom, $\delta
 {\nv}$, locks to the layer normals with $\delta\nv=-\grad_{\perp}u$.
 
  Integrating out the smooth part of $u$ and $\delta \nv$, leads to the constraints:
  \begin{equation}
      \begin{split}
          \grad \cdot {\bf \sigma}&=0,\\
          \jv+ {\bf \sigma}^{\perp}&=0.
      \end{split}
  \end{equation}
  The first equation can be solved in terms of a vector potential ${\bf A}$, $\sigma_i=\epsilon_{ijk}\partial_j A_k, ({\bf{\sigma}}=\grad \times {\bf A})$, which inside the second constraint gives,
  \begin{equation}
    j_i+\epsilon_{ijk}^{\perp}\partial_j A_k=0, 
  \end{equation}
  such that $j_i$ is solved as $j_i=-\epsilon_{ijk}^{\perp}\partial_j A_k$.
  
\begin{figure}[htbp]
  \centering
  \subfigure[] {
    \hspace{0in}\includegraphics[width=0.22\textwidth]{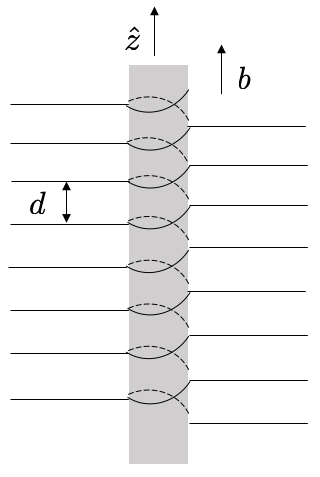}}
    \label{fig:+screwdislocation}
  \hspace{0.32cm}
  \subfigure[]{
    \hspace{0in}\includegraphics[width=0.29\textwidth]{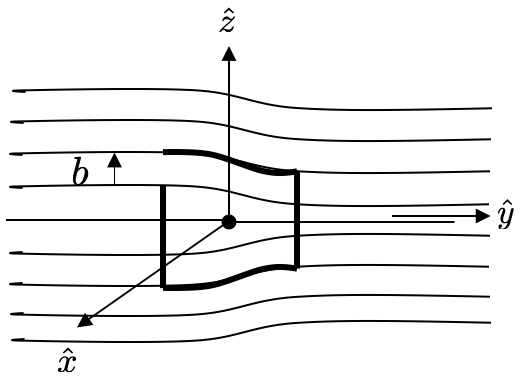}}
    \label{fig:+edgedislocation}
    \caption{Different types of dislocations in a 3D smectic.  (a) A
      screw dislocation with its line core tangent along $\hat{\bf
        z}$. (b) An edge dislocation with its line core
      tangent along $\hat{\bf x}$ axis, i.e., perpendicular to
      $\hat{\bf z}$.}
\label{fig:3Ddefects}
\end{figure}

  Substituting $\sigma_i$ and $j_{ij}$ back into the original Hamiltonian, and integrating by part, lead to
  \begin{equation}
     \begin{split}
         \mathcal{H}_{\text{sm}}^{\text{3d}}=&\frac{1}{2\chi} \left(\grad \times {\bf A}\right)^2-i\epsilon_{ijk} \partial_j A_k \left(\partial_i u+\delta n_i \right)\\-&\frac{1}{2K} \epsilon_{ijk}^{\perp}\partial_j A_k \frac{1}{\grad^2}\epsilon_{imn}^{\perp}\partial_m A_n+i \epsilon_{ijk}^{\perp}\partial_jA_k \delta n_i\\=&\frac{1}{2\chi} \left(\grad \times {\bf A}\right)^2+i A_k \epsilon_{ijk}\partial_j \left(\partial_i u+\delta n_i \right)\\-&\frac{1}{2K} \epsilon_{ijk}^{\perp}\partial_jA_k \frac{1}{\grad^2}\epsilon_{imn}^{\perp}\partial_mA_n-i A_k \epsilon_{ijk}^{\perp}\partial_j \delta n_i\\=&\frac{1}{2\chi} \left(\grad \times {\bf A}\right)^2+i A_k \epsilon_{ijk}\partial_j \partial_i u\\-&\frac{1}{2K} \epsilon_{ijk}^{\perp}\partial_j A_k \frac{1}{\grad^2}\epsilon_{imn}^{\perp}\partial_mA_n+i A_k \left(\epsilon_{ijk}- \epsilon_{ijk}^{\perp}\right)\partial_j \delta n_i\\=&\frac{1}{2\chi} \left(\grad \times {\bf A}\right)^2-\frac{1}{2K} \epsilon_{ijk}^{\perp}\partial_j A_k \frac{1}{\grad^2}\epsilon_{mnk}^{\perp}\partial_m A_n-iA_k b_k,
  \end{split}
\end{equation}
where dislocation density (see Fig.\ref{fig:3Ddefects}) is given by
$b_k=\epsilon_{ijk}\partial_i\partial_j u$.
 
In momentum space, we have
 \begin{equation}
 \begin{split}
     \mathcal{H}_{\text{sm}}^{\text{3d}}=&\frac{1}{2 \chi} \left|\qv \times {\bf A}(\qv)\right|^2+\frac{1}{2K}  \frac{|\left(\qv \times {\bf A}(\qv)\right)_{\perp} |^2}{q^2}\\&-i A_k(\qv) b_k(-\qv)
     \\=&\frac{1}{2 \chi} \left|\qv_{\perp} \times {\bf A}_{\perp}(\qv)\right|^2-i {\bf A}_{\perp}(\qv) \bv_{\perp}(-\qv)\\+&\frac{1}{2}  \left(\frac{1}{\chi}+\frac{1}{Kq^2} \right)|\left(q_z{\bf A}_{\perp}-A_z \qv_{\perp}\right) |^2 -iA_z(\qv)b_z(-\qv),
\end{split}
 \end{equation}
 Functionally integrating out $A_z$, leads to,
 
 \begin{equation}
    \begin{split}
       \mathcal{H}_{\text{sm}}^{\text{3d}}=&\frac{1}{2 \chi} \left[\left|\qv_{\perp} \times {\bf A}_{\perp}(\qv)\right|^2+\left(1+\frac{1}{\lambda^2 q^2}\right) q_z^2|{\bf A}_{\perp}|^2\right]\\&-i {\bf A}_{\perp}(\qv) \bv_{\perp}(-\qv)+\frac{1}{2}\frac{\chi}{ q_{\perp}^2\left[1+1/(\lambda^2q^2)\right]} \left|b_z(\qv)\right|^2\\=&\frac{1}{2\chi}A^{\perp}_i\left[q_{\perp}^2  P_{ij}^{\text{T}} +\left(q_z^2+\frac{q_z^2}{\lambda^2q^2}\right)\left(P_{ij}^{\text{L}}+P_{ij}^{\text{T}} \right)\right]A^{\perp}_j\\&-i {\bf A}_{\perp}(\qv) \bv_{\perp}(-\qv)+\frac{1}{2}\Gamma(\qv) \left|b_z(\qv)\right|^2\\=&\frac{1}{2\chi}A_i^{\perp}(\qv)\left(q^2+\frac{q_z^2}{\lambda^2q^2}\right)P_{ij}^{\text{T}} A_j^{\perp}(-\qv)\\&-i {\bf A}_{\perp}(\qv) \cdot \bv_{\perp}(-\qv)+\frac{1}{2}\Gamma(\qv) \left|b_z(\qv)\right|^2,
   \end{split} 
 \end{equation}
where we have chosen the Coulomb (transverse) gauge $\grad_{\perp} \cdot {\bf A}_{\perp}=0$, i.e., $\qv_{\perp} \cdot {\bf A}_{\perp}=0$, $P^{\text{T}}_{ij}=\delta_{ij}-q_iq_j/q^2$, and, $P^{\text{L}}_{ij}=q_iq_j/q^2$, are transverse and longitudinal projection operators respectively, $\Gamma(\qv)\equiv \chi q_{\perp}^{-2}\left[1+1/(\lambda^2q^2)\right]^{-1}$ is the interaction potential for screw dislocation $b_z$'s, and we have used the fact that, $P_{ij}^{\text{L}}A_j^{\perp}=0$, in the Coulomb gauge. If we integrate out ${\bf A}_{\perp}$ further, we get the dislocation Coulomb gas model, given by,
 \begin{equation}
  \begin{split}
      H_b^{\text{3d}}=&\frac{1}{2}\int_{\qv} \left(\frac{{K}q^2}{q_z^2+\lambda^2 q^4} P_{ij}^{\perp}+2E_b^i \delta_{ij} \right) b_i (\qv) b_j (-\qv)\\&+\frac{1}{2}\int_{\qv}\Gamma(\qv) \left|b_z(\qv)\right|^2\\=&\frac{1}{2}\int_{\qv} \left(\frac{{K} \left[\hat{\bf z} \cdot \qv \times \bv (\qv) \right]^2}{q_z^2+\lambda^2 q_{\perp}^4}+2 E_b^i \bv (\qv) \bv (-\qv) \right),
 \end{split}
 \label{Coulombgas3D}
\end{equation}
where,
$P_{ij}^{\perp}=\left(\delta_{ij}-q_i^{\perp}q_j^{\perp}/q_{\perp}^2\right)(1-\delta_{iz}\delta_{jz})$,
is the transverse projection operator for edge dislocations,
i.e. screw dislocations $b_z-$components projected away, and $E_b^i$'s
are core energies of dislocations \cite{Toner82}. We note that
$\Gamma(\qv)$ is approximately a constant at small $q$, where $q$ is
smaller than $1/\lambda$, and therefore contributes to the core energy
of a screw dislocation, i.e., $E_b^z \to E_b^z+\Gamma(\qv)$. Note that
for simplicity, we have assumed that the lattice spacing between atoms
within the layers is equal to the layer spacing, i.e., $a=d$, such
that we have no factor like `$\frac{d^2}{a^2}$' as in Eq. (\ref{2D
  core}) for the 2D case.
 
Interested in the nature of the nematic to smectic-A transition,
Toner\cite{Toner82} mapped a model of a smectic onto a Coulomb gas of
dislocation loops, which he then transformed into an anisotropic
superconductor in a vector gauge field ${\bf A}$, and analyzed it with
a momentum-shell renormalization group. In the long-wavelength limit
of $q\ll 1/\lambda$, indeed our model reduces to Toner's, with a
generalization that screw dislocations in our model have a finite
interaction.
  
In analogy to what we have done for a 2D smectic, we transform the
Coulomb gas Hamiltonian, Eq.(\ref{Coulombgas3D}) into a classical
gauge theory. The partition function for the dislocation-loop Coulomb
gas on a lattice is given by,
\begin{equation}
  \begin{split}
    Z&=\int \prod_{\rv_n} d{\bf A} (\rv_n)\sum_{\{b_{\rv_n}\}} e^{-H_A
      \left[\{{\bf A}(\rv_n)\}\right]+\sum_{\rv_n} ib_{\rv_n}d \cdot
      {\bf A}(\rv_n)} \\& \cdot e^{-\sum_{\rv_n} E_b^i
      d^2|b^i_{\rv_n}|^2} \delta (\Delta \cdot {\bf b}_{\rv_n})
    \delta(A_z) \delta(\Delta \cdot {\bf A}_{\rv_n})\\&=\int
    \prod_{\rv_n} d{\bf A} (\rv_n) d\varphi({\rv_n})\delta(A_z)
    \delta(\Delta \cdot {\bf A}_{\rv_n}) \\& \cdot e^{-H_A
      \left[\{{\bf A}(\rv_n)\}\right]+\sum_{\rv_n} ib_{\rv_n} \cdot
      \left[d {\bf A}(\rv_n)-\Delta
        \varphi(\rv_n)\right]-E_b^id^2|b^i_{\rv_n}|^2},
\end{split}
\end{equation}
with, 
\begin{equation}
\begin{split}
   H_A \left[\{{\bf A}(\rv_n)\} \right]&=\frac{1}{2}\int \frac{d^3q}{(2\pi)^3} \left(\frac{q_z^2}{{K} q_{\perp}^2}+\frac{q_{\perp}^2}{{\chi}} \right) |{\bf A} (\qv)|^2,
\end{split}
\end{equation}
and we have introduced an auxiliary scalar field $\varphi(\rv)$, such
that integrating out $\varphi$ recovers the constraint, $\Delta \cdot
\bv_{\rv_n}=0$.

After tracing over the dislocation charges $\bv_{\rv_n}$, we obtain
  \begin{equation}
 \begin{split}
   \tilde{H}_{\text{sm}}^{\text{3d}}=H_A-g_{b_i}^{\text{3d}}
   \sum_{\rv_n}\cos \left[\Delta_i \varphi(\rv_n)-d\cdot A_i(\rv_n)
   \right]
\end{split}
\end{equation}
where $g_{b_i}^{\text{3d}}=2 e^{-E_b^i d^2}$, and we have approximated
the resulting Villain potential by its lowest harmonic. In the
continuum limit, it becomes
\begin{equation}
 \begin{split}
    \tilde{H}_{\text{sm}}^{\text{3d}}&=H_A-\tilde{g}_{b_i}^{\text{3d}} \int d^3r \cos \left(\partial_i \varphi-d\cdot A_i \right)\\&=H_A-\int d^3r\left[\frac{1}{2} \left|(\grad-id \cdot {\bf A}) \psi \right|^2+V(|\psi|)\right],
\end{split}
\end{equation}
with,
$\tilde{g}_{b_i}^{\text{3d}}=\frac{g_{b_i}^{\text{3d}}}{a^2d}=\frac{2}{a^2d}
e^{-E_b^i d^2}$, and in the second form we have written it as an
equivalent ``soft-spin'' description in terms of $\psi= |\psi|
e^{i\varphi}$, with the Landau U(1)-invariant potential,
$V\left(|\psi|\right)$, which is equivalent to the first form below
the energy scale of the gapped Higgs-like magnitude degree of freedom,
$|\psi|$. Thus, we reproduce Toner's anisotropic superconductor
model\cite{Toner82} expected to have the same critical
properties as the above dislocation-loop Coulomb gas model.

\section{Summary and conclusion}
In this paper, after a brief review of smectic elasticity, we
developed a coupled U(1) vector gauge theory for a two-dimensional
quantum smectic, where the phonons and orientational Goldstone modes
map onto coupled gauge fields, and topological defects correspond
gauge charges and dipoles.  We discovered that charges (disclinations)
exhibit subdimensional lineon dynamics, restricted to move transverse
to the layer. Motivated by the partial quantum melting of a crystal
into a smectic, and the subsequent smectic-to-nematic transition, we
reproduced the dual description of a quantum smectic by condensing the
one flavor species of dipoles within the generalized Abelian-Higgs
model of a 2D quantum crystal.
 
We also applied this duality to treat a classical smectic liquid
crystal. To this end, we formulated the smectic-to-nematic and
nematic-to-isotropic fluid transitions as a higher-derivative
sine-Gordon model of a 2D classical smectic. Motivated by the
correspondence between a (2+1)D quantum system and a 3D classical
system, we also derived a dual theory for a 3D classical smectic, and
reproduced smectic's dislocation-loop Coulomb gas description for the
nematic-smectic transition, which we then mapped onto an anisotropic
Abelian-Higgs model.
 
We expect this fractonic gauge theory reformulation of smectics will
be useful for further detailed explorations, e.g., subjected to an
external stress and in presence of a substrate. We leave a study of
the true critical behavior (beyond mean-field) of the crystal-smectic
and smectic-nematic transitions using the dual gauge theory for future
studies. The duality analysis in the presence of elastic
nonlinearities also remains a challenging open problem.
  
\label{sec:summary}

\acknowledgments

We acknowledge earlier collaboration with Michael Pretko that
motivated this work. This work was supported by the Simons Investigator
Award from the James Simons Foundation and by the NSF MRSEC grant
DMR-1420736.

\appendix
\section{Crystal-to-supersolid transition }
\label{appendixA}
In the Mott-insulating ``commensurate'' crystal phase associated with
the particle-number conservation symmetry, the Lagrangian density of a
square-lattice quantum-crystal with lattice spacing $d$, for the two
dipole fields $\psi_{\bf p}=\psi_x, \psi_y$, corresponding to the two
minimal dipole species ${\bf p}_x=p \hat{\bf x}, {\bf p}_y=p \hat{\bf
  y}$, plus the Maxwell gauge field part, takes the form
 \begin{equation}
    \begin{split}
       \mathcal{L}_{\text{dis}}=&\sum_{\pv}i\psi^{\dagger}_{\pv} D_0 \psi_{\pv}-\frac{1}{2m_p} \sum_{\pv} |\Pi_{ij}^{\perp {\pv}} D_j \psi_{\pv}|^2- V(\{\psi_k\}) \\&+\mathcal{L}_{\text{M}}^{\text{cr}},
  \end{split}
 \end{equation}
where, $m_p$ is the effective mass of the dipole, $D_0=\partial_t+ip_k A_{0k}$ and $D_j=\partial_j+i p_k A_{jk}$  (i.e., ${\bf D}=\grad +i p_k {\bf A}_k$) are the covariant derivatives\cite{Kumar19}, $V (\{\psi_k\})$ is the U(1)-invariant Ginzburg-Landau potential, $\mathcal{L}_{\mathcal{\text{M}}}^{\text{cr}}$ the Lagrangian density of the Maxwell part, and $\Pi_{ij}^{\perp {\pv}}=\delta_{ij}-\frac{p_i p_j}{p^2}$ is the projection operator since in this Mott-insulating crystal phase, the dipole can only move in the direction perpendicular to {\pv} while the along-dipole climbs are forbidden due to the U(1) particle-number conservation symmetry (`` glide-constraint").

However, as discussed in the main text, the crystal also exhibits scalar non-topological point defects, corresponding to deficiency and excess in atom density, which permits the climb process of the dislocations (See Fig. \ref{fig: Dislocation climb}). Combined with the bosonic statistics of the underlying particles, the quantum crystal can first develop into a super-solid phase (incommensurate crystal), featuring both the crystalline order and the superfluid order. The condensation of vacancies or interstitials in the super-solid phase, therefore, frees these symmetry-forbidden climb events. Therefore, for a complete description, we also need to add the superfluid part of the underlying bosonic particles, $\mathcal{L}_{\text{sf}}$, and the minimal gauge-invariant coupling between the dislocation climb operators and superfluid order parameter ,$\mathcal{L}_{\text{sf-dis}}$,  into the full Lagrangian \cite{Kumar19},
  \begin{eqnarray}
    &&\mathcal{L}_{\text{sf}}=i\Psi^{\dagger}_{\text{sf}} \partial_t \Psi_{\text{sf}}-\frac{1}{2m_s}|\grad \Psi_{\text{sf}}|^2-\mu|\Psi_{\text{sf}}|^2-\frac{U}{2}|\Psi_{\text{sf}}|^4,\;\;\;\;\;\;\;\;\\
    &&\mathcal{L}_{\text{sf-dis}}= \gamma \Psi_{\text{sf}} \sum_{\bf p} \mathcal{O}_{\text{climb}, {\bf p}}+h.c.,
  \end{eqnarray}
where, $\Psi_{\text{sf}}$ is the superfluid order parameter, $\mathcal{O}_{\text{climb}, {\bf p}}=\psi_{\bf p}^{\dagger}({\bf r}+{\bf p}) e^{ip_i A_{ij} p_j} \psi_{\bf p} ({\bf r})$ is the ${\bf p}$-dipole climb operator, and $\gamma$ is the coupling constant\cite{Kumar19}. Fig. \ref{fig:sf-dis. coupling}  shows an example of terms in $\mathcal{L}_{\text{sf-dis}}$.

 The Mott insulator-to-superfluid transition, described by $\mathcal{L}_{\text{sf}}$, occurs at the critical point $\mu=0$, with  $|\Psi_{\text{sf}}|=\sqrt{\frac{|\mu|}{U}}\equiv \sqrt{\rho_s^0}$. Writing $\Psi_{\text{sf}}=\sqrt{\rho_s} e^{i\varphi_s}=\sqrt{\rho_s^0+\delta \rho_s}e^{i\varphi_s}$, the superfluid part becomes.
 \begin{equation}
     \mathcal{L}_{\text{sf}}=-\rho_s \partial_t \varphi_s-\frac{1}{2m_s} \left[ \frac{(\grad \rho_s)^2}{4\rho_s}+\rho_s (\grad \varphi_s)^2\right]-\mu \rho_s-\frac{U}{2}\rho_s^2.
 \end{equation}
 Integrating out the massive magnitude fluctuations $\delta \rho_s$, leads to,
 \begin{equation}
 \begin{split}
     \mathcal{L}_{\text{sf}}&=\frac{1}{2} \partial_t \varphi_s \frac{1}{U-1/(4m_s \rho_s^0) \cdot \grad^2} \partial_t \varphi_s-\frac{\rho_s^0}{2m_s} (\grad \varphi_s)^2\\ &\approx \frac{1}{2U} (\partial_t \varphi_s)^2-\frac{\rho_s^0}{2m_s} (\grad \varphi_s)^2\\&\equiv \frac{\rho_s^0}{2m_s} (\partial_{\mu}\varphi_s)^2,
\end{split}
 \end{equation}
 where, in the second line, we have assumed that $\varphi_s({\bf r})$ varies slowly in space and dropped the term, $\frac{1}{4m_s\rho_s^0}\grad^2$, and in the last line, we have written it as that of a sound mode, in a Lorentz-invariant form for simplicity, with  $\partial_{\mu} \equiv \left(\frac{1}{c_{\text{ph}}}\partial_t, \grad \right)$ and $c_{\text{ph}}=\sqrt{\frac{\rho_s^0 U}{m_s}}$.

Combined, $\mathcal{L}_{\text{dis}}$, with, $\mathcal{L}_{\text{sf}}$ and $\mathcal{L}_{\text{sf-dis}}$, the resulting phase is a supersolid with the spontaneous breaking of both partice number conservation U(1) symmetry and translational symmetry.
Writing $\psi_k=\sqrt{\rho_k}e^{i\varphi_k}$, and integrating out the massive magnitude fluctuations, the effective Lagrangian density for the super-solid phase is given by
\begin{equation}
\begin{split}
    \tilde{\mathcal{L}}_{\text{cr}}=&\frac{\rho_k}{2}  \left(\partial_{t} \varphi_k+p A_{0k} \right)^2-\frac{\rho_k}{2m_{p}}\cos \left[\Pi_{ij}^{\perp k}\left(\partial_j \varphi_k+pA_{jk}\right) \right] \\&-\rho_p^{\parallel} \cos\left[\Pi_{ij}^{\parallel k} \left( \partial_j \varphi_k+p A_{jk}\right)+\varphi_s \right]+\mathcal{L}_{\text{M}}^{\text{cr}}+\mathcal{L}_{\text{sf}}.
\end{split}
\end{equation}

Freezing the superfluid phase by fixing $\varphi_s=0$, and rescaling the longitudinal and transverse gradients, we get (correct to quadratic order in the argument of cosine terms),
\begin{equation}
\begin{split}
    \tilde{\mathcal{L}}_{\text{cr}}=&\frac{\rho_k}{2}  \left(\partial_{t} \varphi_k+p A_{0k} \right)^2-\frac{\rho_k}{2m_{p}}\cos \left(\grad \varphi_k+p {\bf A}_k\right)+\mathcal{L}_{\text{M}}^{\text{cr}}\\=&\frac{1}{2} |\left(\partial_{\mu}+ipA_{\mu, k} \right)\psi_k|^2-V\left(\{\psi_k\}\right)+\mathcal{L}_{\text{M}}^{\text{cr}},
\label{Dual Alternatively}
\end{split}
\end{equation}
where in the second form, we have written it as an equivalent
``soft-spin" description in terms of $\psi_k$, with the landau
U(1)-invariant potential, $V\left(\{\psi_k\}\right)$. Note that the
fugacity $e^{-\tilde{E}_k^b}$ and density $\rho_k$ are related by the
relation: $\frac{m_p}{d^2} e^{-\tilde{E}_k^b}=\rho_k$ to the lowest
order, such that $\frac{\rho_k}{m_p}=2 \tilde{g}_k^b$. Therefore,
Eq.(\ref{Dual Alternatively}) is in the same form as Eq.(\ref{Higgs
  crystal}).

Similar analysis applies for the quantum smectic phase with just one species of dipoles, i.e., $\hat{\bf x}-$dipoles ($\hat{\bf y}-$dislocations), which leads to the full mobility of dipoles (dislocations) in the smectic and the effective Lagrangian density of the super-smectic phase given by,
\begin{equation}
\begin{aligned}
    \tilde{\mathcal{L}}_{\text{sm}}=&\frac{\rho}{2} \left( \partial_t \varphi_x+p A_0\right)^2 -\frac{\rho}{2m_p} \cos\left(\grad \varphi_x+p {\bf A}\right)+\mathcal{L}_{\text{M}}^{\text{sm}}\\=&\frac{1}{2} |\left(\partial_{\mu}+ipA_{\mu} \right)\psi_x|^2-V\left(|\psi_x|\right)+\mathcal{L}_{\text{M}}^{\text{sm}},
\end{aligned}
\end{equation}
which is in the same form as Eq.(\ref{Dual Super-smectic 1}).

\section{From Crystal dual to Smectic dual}
The crystal-smectic partial melting transition corresponds to
condensation of one flavor of the dipole fields, that according to
Fig. \ref{fig:phase transition}, we take to be $\hat{\bf y}$-dipoles,
i.e. $\psi_y \neq 0$. In this $\hat{\bf y}$-dipole condensate, the
vortices in the phase field $\varphi_y$ are suppressed, i.e.,
$\grad \varphi_y$ is small, and thus, we can expand the corresponding
cosine term in (\ref{Dual Alternatively}) to the quadratic order in
its argument, which leads to
\begin{equation}
    \begin{split}
    \tilde{\mathcal{L}}_{\text{cr}}=&\frac{\rho_x}{2}  \left(\partial_{t} \varphi_x+p A_{0x} \right)^2-\frac{\rho_x}{2m_{p}}\cos \left(\grad \varphi_x+p {\bf A}_x\right)+\mathcal{L}_{\text{M}}^{\text{cr}}\\&+\frac{\rho_y}{2}  \left(\partial_{t} \varphi_y+p A_{0y} \right)^2+\frac{1}{2} \cdot \frac{\rho_y}{2m_{p}} \left(\grad \varphi_y+p {\bf A}_y\right)^2.
\end{split}
\end{equation}
With the gauge transformation by choosing
$\chi_y \to \frac{1}{p} \varphi_y$ in Eq.(\ref{gauge transform}),
 \begin{equation}
     A_{iy} \to A_{iy}+\frac{1}{p} \partial_i \varphi_y,\; A_{0y} \to A_{0y}+\frac{1}{p} \partial_t \varphi_y,\; a_y \to a_y+\frac{1}{p}\varphi_y,
 \end{equation}
 we absorb the gradients of the phase $\varphi_y$ into the gauge
 fields $A_{\mu, y}$, and the last two terms in Eq.(B1) become
 quadratic terms of $\hat{\bf y}$-flavor gauge fields, $A_{\mu, y}^2$,
 which make the original massless modes $A_{\mu, y}$ become
 massive. And, $\tilde{\mathcal{L}}_{\text{cr}}$ can be written as
 \begin{equation}
 \begin{split}
    \tilde{\mathcal{L}}_{\text{cr}}=&\frac{\rho_x}{2}  \left(\partial_{t} \varphi_x+p A_{0x} \right)^2-\frac{\rho_x}{2m_{p}}\cos \left(\grad \varphi_x+p {\bf A}_x\right)+\mathcal{L}_{\text{M}}^{\text{eff}},
\end{split}
 \end{equation}
 where the effective Maxwell part $\mathcal{L}_{\text{M}}^{\text{eff}}$ in terms of newly defined $A_{\mu, y}$ is given by
 \begin{equation}
     \mathcal{L}_{\text{M}}^{\text{eff}}=\mathcal{L}_{\text{M}}^{\text{cr}}+\frac{1}{2}\rho_y p^2 A_{0y}^2+\frac{1}{2} \cdot \frac{\rho_y p^2}{2m_{p}} {\bf A}_y^2
 \end{equation}
 
 Therefore, within this condensate phase,  the gauge field components $A_{y,\mu}$ become gapped via the Anderson-Higgs mechanism by coupling to the $\hat{\bf y}$-dipole ($\hat{\bf x}$-dislocation) condensate , and can  be safely integrated out in the low-energy regime with wavelengths of excitations much greater than $\sqrt{\frac{2m_p}{\rho_y p^2}}$ (i.e., penetration length of $A_{\mu,y}$) , which leads to 
 \begin{equation}
 \begin{split}
     \mathcal{L}_{\text{M}}^{\text{eff}}=& \frac{1}{2}\chi {\bf E}_x^2-\frac{1}{2} \left(\grad \times {\bf A}_x \right)^2+\frac{1}{2}K_x e_x^2+\frac{1}{2}K_y e_y^2\\&-\frac{1}{2} \left(\frac{\rho_y p^2}{\rho_y p^2-m_p}\right) \left(\epsilon_{ij}\partial_i a_j+\epsilon_{ix}A_{ix}\right)^2,
 \end{split}
 \end{equation}
  where, the modified bend modulus ${\bf K}=(K_x, K_y)$, with $K^{-1}_x=K^{-1}$ and $K_y^{-1}=K^{-1}-\frac{K^{-1}}{K\rho_y p^2+1}$, becomes anisotropic in this smectic case. 
  In the lowest order approximation, making $\rho_y \to \infty$ (i.e., the condensate is very dense), $K_x=K_y=K$, and $\mathcal{L}_{\text{M}}^{\text{eff}}$ reduces to the Maxwell Lagrangian of the smectic case,
  \begin{equation}
     \begin{split}
     \mathcal{L}_{\text{M}}^{\text{eff}}\approx & \frac{1}{2}\chi {\bf E}_x^2-\frac{1}{2} \left(\grad \times {\bf A}_x \right)^2+\frac{1}{2}K {\bf e}^2-\frac{1}{2} \left(\epsilon_{ij}\partial_i a_j+\epsilon_{ix}A_{ix}\right)^2\\=&\mathcal{L}_{\text{M}}^{\text{sm}}.
 \end{split}
 \end{equation}
 
 Therefore, to lowest order, the crystal's Maxwell Lagrangian reduces to that of a smectic,
\begin{equation}
    \mathcal{L}_{\text{M}}^{\text{cr}} \left(A_{ \mu, x}, A_{\mu, y}=0, a_{\mu} \right) \approx \mathcal{L}_{\text{M}}^{\text{sm}} \left(A_{\mu, x}, a_{\mu} \right),
\end{equation}
which simply corresponds to setting $A_{\mu, y} \approx 0$. And, by
replacing $\rho_x$ simply with $\rho$ in Eq.(B3), the dual Lagrangian
of the crystal reduces to that of the smectic exactly, to the lowest
order,
\begin{equation}
   \tilde{\mathcal{L}}_{\text{cr}} \left(A_{ \mu, x}, A_{\mu, y}=0, a_{\mu}, \psi_x, \psi_y=0 \right) \approx \tilde{\mathcal{L}}_{\text{sm}} \left(A_{\mu, x}, a_{\mu}, \psi_x \right).
\end{equation}

Similarly analysis applies for the further melting with a condensation
of the other, $\hat{\bf x}-$dipoles ($\hat{\bf
  y}-$dislocations). Within this $\psi_x \neq 0$ Higgs phase,
corresponding to a condensation of unbound $\hat{\bf x}-$dipoles
($\hat{\bf y}-$dislocations) in the smectic, the gauge field
components $A_{\mu,x}$ become gapped also, via coupling to the
$\hat{\bf x}-$dipole ($\hat{\bf y}-$dislocation) condensate. This can
be seen easily by making a further gauge transformation with
$\chi_x \to \frac{1}{p} \varphi_x$ in Eq.(\ref{gauge transform}),
 \begin{equation}
     A_{ix} \to A_{ix}+\frac{1}{p} \partial_i \varphi_x, \; A_{0x} \to A_{0x}+\frac{1}{p} \partial_t \varphi_x, \; a_x \to a_x+\frac{1}{p}\varphi_x,
 \end{equation}
 to absorb the gradients of the phase $\varphi_x$ into the gauge fields $A_{\mu, x}$, and expanding the corresponding cosine term in (A6) to the quadratic order in its argument, which leads to,
 \begin{equation}
    \begin{split}
    \tilde{\mathcal{L}}_{\text{sm}}=&\mathcal{L}_{\text{M}}^{\text{sm}}+\frac{1}{2} \rho_x p^2 A_{0x}^2+\frac{1}{2}\cdot \frac{\rho_x p^2}{2m_{p}}{\bf A}_x^2.
\end{split} 
 \end{equation}
 where the last two quadratic terms make the original massless modes $A_{\mu, x}$ become massive now. Integrating out $A_{\mu,x}$ in the low-energy regime with wavelengths of excitations much greater than $\sqrt{\frac{2m_p}{\rho_x p^2}}$ (i.e., penetration length of $A_{\mu,x}$), leads to,
 \begin{equation}
 \begin{split}
     \tilde{\mathcal{L}}_{\text{sm}}=&\frac{1}{2}\left(K^{-1}-\frac{1}{\rho_x p^2+K^{-1}} \right)(\partial_t a_x-\partial_x a_0)^2\\&+\frac{1}{2} K^{-1} \left(\partial_t a_y-\partial_y a_0\right)^2-\frac{1}{2}\left(1+\frac{2m_p}{\rho_xp^2} \right)\left(\grad \times {\bf a}\right)^2\\ \approx &\frac{1}{2} K^{-1}\left(\partial_t {\bf a}-\grad a_0\right)^2-\frac{1}{2}\left(\grad \times {\bf a}\right)^2\\=&\tilde{\mathcal{L}}_{\text{nm}},
\end{split}
 \end{equation}
 in the lowest order approximation, making $\rho_x \to \infty$ (i.e., the condensate is very dense), which is just the dual Lagrangian density of the quantum xy model of a nematic. Therefore,
 \begin{equation}
    \mathcal{L}_{\text{M}}^{\text{sm}} \left(A_{ \mu, x}=0, a_{\mu} \right) \approx \mathcal{L}_{\text{M}}^{\text{nm}} \left(a_{\mu} \right),
\end{equation}
to the lowest order.

\section{3D smectic elasticity}
In the main text, we have given the elasticity of a d-dimensional
smectic in terms of the layer displacement $u$ only, given by
Eq.(\ref{H_sm, d}). For a 3D smectic, we just set $d=3$. Here, we
formulate the elasticity of a 3D smectic in terms of the displacement
$u$ and the Frank director $\nv$ simultaneously. The elastic energy
will not change, if all layers of molecules are rotated together
rigidly. However, there will be an energy cost if the orientation
directions of molecules, represented by Frank director $\nv$, are
rotated away from their equilibrium local orientation, normal to the
layers. The elastic energy density of a 3D smectic, with its layers
along $xy$ plane, is given by
 \begin{equation}
  \begin{split}
      \mathcal{H}_{\text{sm}}^{\text{3d}}=&\frac{1}{2}\chi \left(\partial_z u \right)^2+\frac{1}{2}K \left(\grad_{\perp} u+\delta \nv\right)^2+\frac{1}{2}K_1 \left(\grad \cdot {\bf n} \right)^2\\&+\frac{1}{2}K_2 \left[\nv \cdot (\grad \times \nv) \right]^2+\frac{1}{2} K_3 \left[\nv \times (\grad \times \nv) \right]^2,
 \label{3D smectic}
 \end{split}
 \end{equation}
where $u$ is the layer displacement, $\delta \nv=\nv-\hat{z}=\left(\delta n_x, \delta n_y, 0 \right)$ represents the layer orientation degree of freedom, and the last three terms represents the slay, twist and bend distortions of the director respectively, with three independent, corresponding elastic constants $K_1$, $K_2$ and $K_3$. To linear order in $\delta \nv$,
\begin{equation}
  \begin{split}
      \mathcal{H}_{\text{sm}}^{\text{3d}}\simeq&\frac{1}{2}\chi \left(\partial_z u \right)^2+\frac{1}{2}K \left(\grad_{\perp} u+\delta \nv\right)^2+\frac{1}{2}K_1 \left(\grad \cdot {\bf n} \right)^2\\&+\frac{1}{2}K_2 \left[{\bf z} \cdot (\grad \times \nv) \right]^2+\frac{1}{2} K_3 \left[{\bf z} \times (\grad \times \nv) \right]^2\\=&\frac{1}{2}\chi q_z^2|u(\qv)|^2+\frac{1}{2}K \left|\qv_{\perp} u(\qv)+\delta \nv(\qv)\right|^2\\+&\frac{1}{2}K_1 \delta n_i P_{ij}^{\text{L}}\delta n_j+\frac{1}{2}K_2 \delta n_i P_{ij,z}^{\text{T}}\delta n_j+\frac{1}{2} K_3 \delta n_i P_{ij,\perp}^{\text{T}}\delta n_j,
 \end{split}
 \end{equation}
where, $P_{ij}^{\text{L}}=q_i q_j/q^2$ is the longitudinal projector, $P_{ij,z}^{\text{T}}=\left(\delta_{ij}-q_i^{\perp}q_j^{\perp}/q^2\right)(1-\delta_{iz}\delta_{jz})$ and $P_{ij,\perp}^{\text{T}}=\left(\delta_{ij}-q^2_{\perp}/q^2\right)\left(\delta_{ix}\delta_{jx}+\delta_{iy}\delta_{jy}\right)$ are the transverse-to-$\hat{\bf z}$ and transverse-to-layer projector, respectively, and in the second line, we have transformed into the momentum space. For long-wavelength
limit, with the wave number $q\ll\sqrt{\chi/K_2}, \sqrt{\chi/K_3}$, we can integrate out the higher-energy terms, which sets $\delta \nv=\grad_{\perp}u$ and reduces $\mathcal{H}_{\text{sm}}^{\text{3d}}$ into the form,
\begin{equation}
   \begin{split}
      \mathcal{H}_{\text{sm}}^{\text{3d}}=&\frac{1}{2}\chi \left(\partial_z u \right)^2+\frac{1}{2}K_1 \left(\grad_{\perp}^2 u\right)^2,
 \label{3D smectic with u only}
 \end{split} 
\end{equation}
matching exactly with the standard form of the elastic energy of a 3D smectic.
In Section \ref{sec:Classical limit}, for a simple analysis with losing much qualitative physics, we have set $K=\chi$, and made the isotropic elasticity approximation with $K_1=K_2=K_3$, replacing them by $K$ for simplicity.

Below, we give a briefly discussion of dislocations and their energies
in a 3D smectic, based on $\mathcal{H}_{\text{sm}}^{\text{3d}}$ given
by Eq.(\ref{3D smectic with u only}). A more detailed discussion based
on $\mathcal{H}_{\text{sm}}^{\text{3d}}$ given by Eq.(\ref{3D
  smectic}) can be found standard textbooks\cite{Chaikin2000}.  For a
single positive screw dislocation with its line core, located in the
origin of $xy-$plane, in the $\hat{\bf z}$ direction, as shown in
Fig. \ref{fig:3Ddefects}(a) , from the Eq.(\ref{layer locations}),
that determines the positions of the layer planes, we get the layer
displacement given by,
\begin{equation}
  u_{\text{scrw}}(\rv)=\frac{d}{2\pi} \tan^{-1} \frac{y}{x}, 
\end{equation}
taking place in the $xy$ plane only, and then,
\begin{equation}
\begin{aligned}
   \partial_zu_{\text{screw}}&=0,\\ \grad_{\perp}u_{\text{screw}}&=\frac{d}{2\pi}\frac{-y \hat{\bf x}+x \hat{\bf y}}{x^2+y^2}=-\frac{d}{2\pi}\frac{\hat{\bf \varphi}}{r_{\perp}},
\end{aligned}
\end{equation}
shuch that,
\begin{equation}
    \grad_{\perp}^2 u_{\text{screw}}=0,
\end{equation}
and therefore, the energy of a single screw dislocation in a smectic is $0$.

For a single positive edge dislocation with its line core perpendicular to the $\hat{z}$ direction, say in the $\hat{x}$ direction, as shown in Fig. \ref{fig:3Ddefects}(b), the layer displacement is then given by,
\begin{equation}
    u_{\text{edge}}(\rv)=\frac{d}{2\pi} \tan^{-1}\frac{z}{y},
\end{equation}
and then,
\begin{equation}
\begin{aligned}
    \partial_z u_{\text{edge}}&=\frac{y}{y^2+z^2},\\
    \grad_{\perp} u_{\text{edge}}&=-\frac{z\hat{\bf y}}{y^2+z^2},
\end{aligned}
\end{equation}
such that,
\begin{equation}
    \grad_{\perp}^2 u_{\text{edge}}=\frac{2yz}{(y^2+z^2)^2},
\end{equation}
and, the energy of a single edge dislocation in a smectic can be shown to be divergent as the length scale of the system, after integrating (C3) over space.


\begin{thebibliography}{99}
\bibitem{ProstDeGennes}P. G. de Gennes and J. Prost, \emph{The Physics
    of Liquid Crystals, 2nd Edition}. Clarendon Press, Oxford (1993).
\bibitem{Chaikin2000} P. M. Chaikin and T. C. Lubensky, \emph{Principles of Condensed Matter Physics}. Cambridge University Press (2000). 
\bibitem{FF} P. Fulde and R.A. Ferrell, \emph{Superconductivity in a Strong Spin-Exchange Field}, Phys. Rev. {\bf 135}, A550 (1964). 
\bibitem{LO} A.I. Larkin and Yu. N. Ovchinnikov, \emph{Nonuniform state of superconductors}, Sov. Phys. JETP {\bf 20}, 762 (1965). 
\bibitem{LR_VishwanathPRL} L. Radzihovsky and A. Vishwanath,
  \emph{Quantum Liquid Crystals in an Imbalanced Fermi Gas:
    Fluctuations and Fractional Vortices in Larkin-Ovchinnikov
    States}, Phys. Rev. Lett. {\bf 103}, 010404, (2009).
\bibitem{LRpra}L. Radzihovsky, \emph{Fluctuations and phase transitions in Larkin-Ovchinnikov liquid-crystal states of a population-imbalanced resonant Fermi gas}, Phys. Rev. A. {\bf 84}, 023611 (2011). 
\bibitem{HuiZhai}Hui Zhai, \emph{Degenerate quantum gases with spin-orbit coupling: a review}, Rep. Prog. Phys. {\bf 78}, 026001 (2015).
\bibitem{LR_ChoiPRL}L. Radzihovsky and S. Choi, \emph{p-Wave Resonant Bose Gas: A Finite-Momentum Spinor Superfluid}, Phys. Rev. Lett. {\bf 103}, 095302 (2009).

\bibitem{EisensteinQSm}M. P. Lilly, K. B. Cooper, J. P. Eisenstein,
  L. N. Pfeiffer, and K. W. West, \emph{Evidence for an Anisotropic
    State of Two-Dimensional Electrons in High Landau Levels},
  Phys. Rev. Lett. {\bf 82}, 394 (1999).
\bibitem{CsathyARCMP} K. A. Schreiber and G. A. Cs\'athy,
  \emph{Competition of pairing and nematicity in the two-dimensional
    electron gas} Ann. Rev. Cond. Mat. Phys. {\bf 11}, 17 (2020).
\bibitem{Fogler}A. A. Koulakov, M. M. Fogler, and B. I. Shklovskii, \emph{Charge Density Wave in Two-Dimensional Electron Liquid in Weak Magnetic Field}, Phys. Rev. Lett. {\bf 76}, 499 (1996).
\bibitem{Moessner}R. Moessner and J. T. Chalker, \emph{Exact results for interacting electrons in high Landau levels}, Phys. Rev. B {\bf 54}, 5006 (1996).
\bibitem{FisherMacdonald}Emiliano Papa, John Schliemann, A. H. MacDonald, and Matthew P. A. Fisher, \emph{Quantum theory of bilayer quantum Hall smectics}, Phys. Rev. B {\bf 67}, 115330 (2003).
\bibitem{LR_Dorsey}L. Radzihovsky and A. T. Dorsey, \emph{Theory of
    Quantum Hall Nematics}, Phys. Rev. Lett. {\bf 88}, 216802 (2002).
  
\bibitem{TranquadaStripes} J. M. Tranquada, et. el., {\em Coexistence
    of, and competition between, superconductivity and charge-stripe
    order in LaNdSrCuO}, Phys. Rev. Lett. {\bf 78}, 338 (1997).
\bibitem{KivelsonStripes} S. A. Kivelson, E. Fradkin, V. J. Emery, {\em
    Electronic liquid crystal phases of a doped Mott insulator},
  Nature {\bf 393}, 550-553 (1998).

\bibitem{HalperinOstlund}S. Ostlund and B. I. Halperin, \emph{Dislocation-mediated melting of anisotropic layers}, Phys. Rev. B {\bf 23}, 335 (1981).

\bibitem{KT} J. M. Kosterlitz and D. J. Thouless, \emph{Ordering, metastability and phase transitions in two-dimensional systems}, J. Phys. C {\bf 6}, 1181 (1972).

\bibitem{Landau37} L. D. Landau, \emph{On the theory of phase transition}, Phys. Z. Sowjetunion \uppercase\expandafter{\romannumeral2}, 26 (1937).  [Eng. trans.: Collected papers of L.D. Landau, ed. D. ter Haar (Gordon and Breach, New York, 1965), pp. 193-217].
\bibitem{Peierls36} R. Peierls, Helv. Phys. Acta., Suppl. \uppercase\expandafter{\romannumeral2} {\bf 7}, 81 (1936).

\bibitem{TonerNelson} John Toner and David R. Nelson, \emph{Smectic,
    cholesteric, and Rayleigh-Benard order in two dimensions},
  Phys. Rev. B {\bf 23}, 316 (1981).

 \bibitem{Grinstein82} G. Grinstein and Robert A. Pelcovits, \emph{Nonlinear elastic theory of smectic liquid crystals}, Phys. Rev. A {\bf 26}, 915 (1982).
 
 \bibitem{LR2011PRE} L. Radzihovsky and T. C. Lubensky, \emph{Nonlinear smectic elasticity of helical state in cholesteric liquid crystals and helimagnets}, Phys. Rev. E {\bf 83}, 051701 (2011).
  
\bibitem{Smecticgauge} L. Radzihovsky, \emph{Quantum smectic gauge 
    theory}, arXiv:2009.06632 [cond-mat.str-el], Phys. Rev. Lett. {\bf 
    125}, 267601 (2020). 
  
\bibitem{Chamon05}C. Chamon, \emph{Quantum Glassiness in Strongly Correlated Clean Systems: An Example of Topological Overprotection},
  Phys. Rev. Lett. {\bf 94}, 040402 (2005).
\bibitem{Bravyi11}S. Bravyi, B. Leemhuis, and B. M. Terhal,
  \emph{Topological order in an exactly solvable 3D spin model},
  Ann. Phys. (Amsterdam) {\bf 326}, 839 (2011).
\bibitem{Haah11} J. Haah, \emph{Local stabilizer codes in three dimensions without string logical operators}, Phys. Rev. A {\bf 83}, 042330 (2011).
\bibitem{Castelnovo12}C. Castelnovo and C. Chamon, \emph{Topological quantum glassiness}, Philos. Mag. {\bf 92}, 304 (2012).
\bibitem{Yoshida13} B. Yoshida, \emph{Exotic topological order in fractal spin liquids},
Phys. Rev. B {\bf 88}, 125122 (2013).
\bibitem{Bravyi13}S. Bravyi and J. Haah, \emph{Quantum Self-Correction in the 3D
Cubic Code Model}, Phys. Rev. Lett. {\bf 111}, 200501 (2013).
\bibitem{Vijay15}S. Vijay, J. Haah, and L. Fu, \emph{A new kind of topological quantum order: A dimensional hierarchy of quasiparticles built from stationary excitations}, Phys. Rev. B {\bf 92}, 235136
(2015).
\bibitem{Vijay16} S. Vijay, J. Haah, and L. Fu, \emph{ Fracton topological order,
generalized lattice gauge theory and duality}, Phys. Rev. B
{\bf 94}, 235157 (2016).

\bibitem{QiAOP2020} M. Qi, L. Radzihovsky, and M. Hermele, \emph{Fracton phases via exotic higher-form symmetry-breaking}, Annals of Physics, 168360 (2020). 

\bibitem{HNreview} R. M. Nandkishore and M. Hermele, \emph{Fractons}, Annu. Rev. Condens. Matter Phys. {\bf 10}, 295 (2019).
\bibitem{AbhinavPremReview}
Abhinav Prem, Michael Pretko, and Rahul M. Nandkishore, \emph{Emergent
  phases of fractonic matter}, Phys. Rev. B {\bf 97}, 085116 (2018).

\bibitem{Pretko1703} M. Pretko, \emph{Subdimensional particle structure of higher rank
U(1) spin liquids}, Phys. Rev. B {\bf 95}, 115139 (2017). 
\bibitem{Pretko1707} M. Pretko, \emph{Generalized electromagnetism of subdimen-
    sional particles}, Phys. Rev. B {\bf 96}, 035119 (2017).
\bibitem{Slagle} K. Slagle and Y. B. Kim, \emph{Fracton topological order from nearest-neighbor two-spin interactions and dualities}, Phys. Rev. B {\bf 96}, 165106 (2017). 
\bibitem{RadzihovskyConjectureDuality16} L. Radzihovsky, unpublished (2016).
\bibitem{dasgupta} C. Dasgupta and B. I. Halperin, \emph{Phase
    transition in a lattice model of superconductivity},
  Phys. Rev. Lett. 47, 1556 (1981).
\bibitem{fisher} M. P. A. Fisher and D. H. Lee, \emph{Correspondence
    between two-dimensional bosons and a bulk superconductor in a
    magnetic ﬁeld},  Phys. Rev. B 39, 2756 (1989).
\bibitem{PretkoLRdualityPRL2018}M.  Pretko  and  L.  Radzihovsky,
  \emph{Fracton-elasticity duality}, Phys.  Rev.  Lett. {\bf 120},
  195301 (2018).
  \bibitem{Zaanen2017} J. Beekman, J. Nissinen, K. Wu, K. Liu, 
  R.-J. Slager, Z. Nussinov, V. Cvetkovic, and J. Zaanen, \emph{Dual gauge field theory of quantum liquid crystals in two dimensions}, 
  Phys. Rep. {\bf 683}, 1 (2017). 

  \bibitem{GromovDualityPRL2019}Andrey Gromov, \emph{Chiral
      Topological Elasticity and Fracton Order}, Phys. Rev. Lett. {\bf
      122}, 076403 (2019).
    
\bibitem{PretkoLRsymmetryEnrichedPRL2018}M.  Pretko  and  L.  Radzihovsky, \emph{Symmetry-Enriched Fracton Phases from Supersolid Duality}, Phys.  Rev.  Lett. {\bf 121}, 235301 (2018).
\bibitem{PretkoZhaiLRdualityPRB}M Pretko, Z Zhai and L Radzihovsky, \emph{Crystal-to-fracton tensor gauge theory dualities},
Phys. Rev. B {\bf 100}, 134113 (2019).
\bibitem{Kumar19}Ajesh Kumar and Andrew C. Potter, \emph{Symmetry-enforced fractonicity and two-dimensional quantum crystal melting},
  Phys. Rev. B {\bf 100}, 045119 (2019).

\bibitem{RadzihovskyHermeleVectorGaugePRL2020}L. Radzihovsky and M. Hermele, \emph{Fractons from vector gauge theory}, Phys. Rev. Lett. {\bf 124}, 050402 (2020). 


\bibitem{Stein78} D. L. Stein, \emph{Kosterlitz-Thouless phase
    transitions in two-dimensional liquid crystals}, Phys. Rev. B {\bf
    18}, 2397 (1978); D. R. Nelson and J. M. Kosterlitz,
  \emph{Universal Jump in the Superfluid Density of Two-Dimensional
    Superfluids}, Phys. Rev. Lett. {\bf 39}, 1201 (1977).

\bibitem{deGennes72}P. G. DeGennes, \emph{An analogy between superconductors and smectics A}, Solid State Commun. {\bf 10}, 753 (1972).

\bibitem{HalperinMa} B. I. Halperin, T. C. Lubensky, and S. K. Ma, \emph{First-Order Phase Transitions in Superconductors and Smectic-
A Liquid Crystals}, Phys. Rev. Lett. {\bf 32}, 292 (1974).

\bibitem{Helfrich78} W. Helfrich, \emph{Defect model of the smectic A-nematic phase transition}, J. Phys. (Paris) {\bf 39}, 1199 (1978).
\bibitem{NelsonToner} D. R. Nelson and J. Toner, \emph{Bond-orientational order, dislocation loops, and melting of solids and smectic-A liquid crystals}, Phys. Rev. B {\bf 24}, 363 (1981).

\bibitem{Lubensky81} T. C. Lubensky, S. G. Dunn, and Joel Isaacson, \emph{Gauge Transformations and the Nematic to Smectic-A Transition}, Phys. Rev.
Lett. {\bf 47}, 1609 (1981).

\bibitem{Grinstein86}G. Grinstein, T. C. Lubensky, and J. Toner, \emph{Defect-mediated melting and new phases in three-dimensional systems with a single soft direction},
Phys. Rev. B {\bf 33}, 3306 (1986).

\bibitem{Toner82} J. Toner, \emph{Renormalization-group treatment of the dislocation loop model of the smectic-A-nematic transition}, Phys. Rev. B {\bf 26}, 462 (1982).

\bibitem{MarchettiRadzihovsky}  M. C. Marchetti and L. Radzihovsky, \emph{Interstitials, vacancies 
    and dislocations in flux-line lattices: A theory of vortex 
    crystals, supersolids and liquids}. Phys. Rev. B 59, 12001 (1999), 
  arXiv:cond-mat/9811193v2. 
  
\bibitem{Nelson79} D. R. Nelson and B. I. Halperin, \emph{Dislocation-mediated melting in two dimensions}, Phys. Rev. B {\bf 19}, 2457 (1979).
\bibitem{Halperin79} B. I. Halperin, \emph{Superfluidity, melting and liquid-crystal phases in two dimensions}, in Proceeding of Kyoto Summer Institute 1979- Physics of Low Dimensional Systems, edited by Y. Nagaoka and S. Hikami (Publications Office, Progress of Theoretical Physics, Kyoto, 1979). 
\bibitem{sineGordon} J. V. Jos\'e, L. P. Kadanoff, S. Kirkpatrick, and D. R. Nelson, \emph{Renormalization, vortices, and symmetry-breaking perturbations in the two-dimensional planar model}, Phys. Rev. B {\bf 16}, 1217 (1977).

\bibitem{Zhai19}Z. Zhai and L. Radzihovsky, \emph{Two-dimensional melting via sine-Gordon duality}, Phys. Rev. B {\bf 100}, 094105 (2019). 

 \end{thebibliography}
\end{document}